\documentclass[a4paper,10pt]{article}
\pdfoutput=1
\usepackage{geometry}
\usepackage{graphicx}
\usepackage{epsf}
\usepackage{amsmath}
\usepackage{amssymb}

\usepackage{cite}
\usepackage{multirow,tabularx}
\usepackage{appendix}
\usepackage{tikz,parskip}
\usepackage{graphicx,amsmath,amsfonts,amssymb,amsthm,euscript,braket,xcolor,wasysym}
\newcommand{\be}{\begin{equation}}
\newcommand{\ee}{\end{equation}}

\newcommand{\Rmnum}[1]{\expandafter\@slowromancap\romannumeral #1@}
\newcommand{\bea}{\begin{eqnarray}}
\newcommand{\eea}{\end{eqnarray}}

\numberwithin{equation}{section}

\newcommand*\circled[1]{\tikz[baseline=(char.base)]{
            \node[shape=circle,draw,inner sep=2pt] (char) {#1};}}

\begin{document}

\title{\bf Anisotropic string tensions and inversely magnetic catalyzed deconfinement from a dynamical AdS/QCD model}

\author{\textbf{Hardik Bohra$^{a}$}\thanks{hardikbohra.nitr@gmail.com}, \textbf{David Dudal$^{b,c}$}\thanks{david.dudal@kuleuven.be},
\textbf{Ali Hajilou$^{d,b}$}\thanks{ali.hajilou@student.kuleuven.be}, \textbf{Subhash Mahapatra$^{a}$}\thanks{mahapatrasub@nitrkl.ac.in}
 \\\\
 \textit{{\small $^a$ Department of Physics and Astronomy, National Institute of Technology Rourkela, Rourkela - 769008, India}}\\
\textit{{\small $^b$ KU Leuven Campus Kortrijk -- Kulak, Department of Physics, Etienne Sabbelaan 53 bus 7657,}}\\
\textit{{\small 8500 Kortrijk, Belgium}}\\
\textit{{\small $^c$  Ghent University, Department of Physics and Astronomy, Krijgslaan 281-S9, 9000 Gent, Belgium}}\\
\textit{{\small $^d$ Department of Physics, Shahid Beheshti University G.C., Evin, Tehran 19839, Iran}}}
\date{}


\maketitle
\abstract{We extend previous work on dynamical AdS/QCD models by introducing an extra ingredient under the form of a background magnetic field, this to gain insight into the influence such field can have on crucial QCD observables. Therefore, we construct a closed form analytic solution to an Einstein-Maxwell-dilaton system with a magnetic field. We specifically focus on the deconfinement transition, reporting inverse magnetic catalysis, and on the string tension, reporting a weaker/stronger confinement along/perpendicular to the magnetic field. The latter, being of importance to potential modelling of heavy quarkonia, is in qualitative agreement with lattice findings.}

\section{Introduction}
Quantum chromodynamics (QCD) is the quantum field theory of strong interactions capable of describing sub-atomics particles such as quarks and gluons. Two of its main characteristic features are color confinement and chiral symmetry breaking \cite{Roberts:1994dr}. It is by now well known that QCD at low temperature and chemical potential exhibits confinement and chiral symmetry breaking whereas at high temperature and chemical potential it undergoes a phase transition to a chiral symmetry restored phase where deconfinement also sets in. Understanding the complete phase diagram of QCD in the parameter space of temperature, chemical potential etc. is a challenging task, and is of utmost importance in high energy physics. Indeed, the investigation of the QCD phase diagram and the search of new phases of matter are of great relevance, attracting worldwide attention, be it from the experimental, lattice or theoretical communities \cite{Stephanov:2007fk}.

Recent experiments with relativistic heavy ion collisions have suggested the possibility of new parameters in the QCD phase diagram. In particular, it is expected that a very strong magnetic field $eB \sim 0.3~\text{GeV}^2$, which is created in the early stages of noncentral relativistic heavy ion collision \cite{Skokov:2009qp,Bzdak:2011yy,Voronyuk:2011jd,Deng:2012pc,Tuchin:2013ie}, might have important consequences on the QCD phase diagram \cite{Miransky:2015ava,Kharzeev:2012ph,DElia:2010abb}. The extremely large magnetic field rapidly decays after the collision, however, it is expected that it still remains sufficiently large at the time when the quark-gluon plasma (QGP) forms \cite{McLerran:2013hla,Tuchin:2013apa}, hence it can affect the QCD matter near the deconfinement transition temperature \cite{DElia:2010abb}. This expectation has led to an intense investigation of QCD in the presence of a background magnetic field. Because of its many interesting properties and phenomenological relevance for e.g.~the chiral magnetic effect \cite{Fukushima:2008xe,Kharzeev:2007jp}, (inverse) magnetic catalysis \cite{Miransky:2002rp,Gatto:2010pt,Mizher:2010zb,Osipov:2007je,Kashiwa:2011js,Alexandre:2000yf,Fraga:2008um,Fukushima:2012xw,Semenoff:1999xv,Shovkovy:2012zn,Bali:2011qj,Bali:2012zg,
Ilgenfritz:2013ara,Bruckmann:2013oba,Fukushima:2012kc,Ferreira:2014kpa,Mueller:2015fka,Bali:2013esa,Fraga:2012fs,Ayala:2014iba,Ayala:2014gwa,Fraga:2012ev}, early universe physics \cite{Grasso:2000wj,Vachaspati:1991nm}, dense neutron stars \cite{Duncan:1992hi} etc., the area of magnetised QCD drew much attention in recent years. For detailed reviews on these subjects, see for example \cite{Miransky:2015ava,Kharzeev:2012ph}.

It was expected from the work of \cite{Gusynin:1994xp,Gusynin:1994re} that the magnetic field has a constructive effect on the quark condensate and the deconfinement transition temperature, a phenomenon commonly termed as the magnetic catalysis. Further investigations, both from lattice simulations and from theoretical models based on weak coupling approximations, had confirmed these results \cite{Miransky:2002rp,Gatto:2010pt,Mizher:2010zb,Osipov:2007je,Kashiwa:2011js,Alexandre:2000yf,Fraga:2008um,Fukushima:2012xw,Semenoff:1999xv,Shovkovy:2012zn}. However, it came as a big surprise when state of the art lattice calculations instead found inverse magnetic catalysis, i.e.~the magnetic field was found to facilitate the destruction of the quark condensate and thence decreased the transition temperature \cite{Bali:2011qj,Bali:2012zg,Ilgenfritz:2013ara}. Subsequently, several interesting physical scenarios, though not entirely satisfactory, were suggested to explain the reason behind the inverse magnetic catalysis, see for example \cite{Bruckmann:2013oba,Fukushima:2012kc,Ferreira:2014kpa,Mueller:2015fka,Bali:2013esa,Fraga:2012fs,Ayala:2014iba,Ayala:2014gwa,Fraga:2012ev}.

It is widely expected that the inverse magnetic catalysis behaviour mainly results from the strongly coupled dynamics around the transition temperature, as most perturbative QCD calculations and other effective models instead suggest magnetic catalysis. Since the dominant physics near the deconfinement and chiral phase transition is non-perturbative, it is therefore more appropriate to study the magnetic field effects in QCD using techniques that are more reliable at strong coupling. Here the idea of gauge-gravity duality, apart from the usual lattice calculations, appears as a natural candidate \cite{Maldacena:1997re,Gubser:1998bc,Witten:1998qj}. The idea that certain strongly coupled field theories without any gravitational degrees of freedom can be mapped to classical Einstein gravity via the gauge-gravity duality provides an elegant method by which the strongly coupled regime of QCD can be probed. Indeed by now, after certain modifications of the original model of the gauge-gravity duality, many characteristic features of QCD have been reproduced from holography, and in some cases, new and interesting predictions have also been made, for a recent review see \cite{Gursoy:2010fj}.

In recent years, the effect of a background magnetic field on the chiral condensate and deconfinement transition has also been discussed holographically. This question has been addressed in both top-down as well as bottom-up models of holographic QCD \cite{Johnson:2008vna,Callebaut:2013ria,Alam:2012fw,Preis:2010cq,Filev:2010pm,Dudal:2015wfn,Mamo:2015dea,Li:2016gfn,Evans:2016jzo,Bolognesi:2011un,
Ballon-Bayona:2017dvv,Rodrigues:2017cha,Rodrigues:2017iqi,McInnes:2015kec,Gursoy:2017wzz,Gursoy:2016ofp,Dudal:2016axr,Dudal:2018rki,Rodrigues:2018pep}. The top-down models, ---unfortunately, not entirely appropriate to describe QCD-like physics in the first place, but more satisfactory as far as the correctness and validness of the duality is concerned---predict magnetic catalysis \cite{Johnson:2008vna}. On the other hand, some phenomenological bottom-up holographic soft and hard wall QCD models do predict inverse magnetic catalysis for the deconfinement transition, however, they continue to show magnetic catalysis behaviour for the chiral transition \cite{Dudal:2015wfn,Bolognesi:2011un}. Moreover, these soft and hard wall models do not always solve the gravity equations explicitly and in most cases the running of the coupling constant (or the dual dilaton field) is introduced by hand in an ad hoc way into the Einstein-Hilbert action. In recent times, more advanced phenomenological bottom-up holographic QCD models, which correctly solve the gravity equations, have been constructed that display inverse magnetic catalysis. In $2+1$ dimensions, sensible gravity solutions displaying inverse magnetic catalysis have been found in \cite{Rodrigues:2017cha,Rodrigues:2017iqi,Rodrigues:2018pep}, while in $3+1$, \cite{Gursoy:2016ofp,Gursoy:2017wzz} displayed the possibility of inverse catalysis in the deconfinement as well as in the chiral sector, depending on the value of a new parameter $c$ which can influence AdS/QCD at vanishing magnetic field as well. The specific r\^{o}le and influence of this parameter $c$ is to the best of our knowledge an interesting open question.

Another interesting inherently non-perturbative QCD quantity, with direct observable consequences, is the string tension between heavy quarks. This is relevant to understanding the binding (and consequent melting at higher temperatures) of heavy quark states such as charmonium, in particular when relying on potential modelling \cite{Bonati:2015dka,Alford:2013jva}. Original lattice data of \cite{Bonati:2014ksa,Bonati:2016kxj} predicts an increase, respectively decrease, in the string tension perpendicular, respectively, parallel to the quark-antiquark orientation. Supporting evidence for such scenario came in from modelling the non-perturbative QCD vacuum in a specific way \cite{Simonov:2015yka}. In \cite{Chernodub:2010bi}, a semiclassical reasoning was provided to argue against QCD string breaking in the perpendicular direction for sufficiently large magnetic field. To our knowledge, the magnetic field induced anisotropies in the string tension are not that well explored from a holographic viewpoint (a general discussion can be found in \cite{Gursoy:2018ydr}) and we want to bridge this gap\footnote{Although the anisotropic effects of the background magnetic field on the probe quark-antiquark free energy have been discussed previously in \cite{Rougemont:2014efa}, however, it did not provide any concrete result on the string tension, which is not surprising given the non-confining $\mathcal{N}=4$ SYM setup of the latter work.}. Of course, specific effects that anisotropy can have in a holographic context, have been explored, see e.g.~\cite{Mateos:2011ix,Mateos:2011tv,Giataganas:2017koz,Giataganas:2012zy,Arefeva:2018cli}.

One of the main problems that have hindered the construction of a genuine phenomenological holographic QCD model with a background magnetic field is the difficulty to find a dilaton backreacted magnetised AdS solution. This will require a simultaneous solution of the Einstein-Maxwell-dilaton system with a non-trivial and consistent profile for the dilaton field. Indeed, a magnetic field embedded Einstein-Maxwell-dilaton gravity system corresponds to a few second-order non-linear coupled differential equations. Closed-form analytical solutions are rather difficult to be found.  In this work, using the potential reconstruction method \cite{He:2013qq,Yang:2015aia,Dudal:2017max,Arefeva:2018hyo,Dudal:2018ztm,Mahapatra:2018gig,Mahapatra:2019uql,Li:2011hp,Cai:2012xh,Alanen:2009xs}, we will remedy this problem and find a complete solution to the Einstein-Maxwell-dilaton gravity system, containing both the magnetic field as well as the running dilaton. In particular, we will show that a consistent solution to the Einstein-Maxwell-dilaton system can be found in terms of a single scale function $A(z)$ (see eqs. (\ref{Atsol}), (\ref{gsol}), (\ref{f2sol}), (\ref{phisol}) and (\ref{Vsol})). This scale function will be further chosen by taking inputs from real QCD and by matching holographic QCD results with real QCD with vanishing magnetic field. Moreover, in addition to the finite temperature and magnetic field, we
extend our model to include the chemical potential as well. This is desirable as computations with finite chemical potential are currently very
challenging for lattice techniques due to the well-known sign problem in Euclidean space-time.

\section{Einstein-Maxwell-dilaton gravity with a magnetic field}
In order to construct a magnetised holographic QCD model with running dilaton and chemical potential, we consider a five dimensional Einstein-Maxwell-dilaton (EMD) gravity system with two Maxwell fields,
\begin{eqnarray}
S_{EM} =  -\frac{1}{16 \pi G_5} \int \mathrm{d^5}x \sqrt{-g}  \ \left[R - \frac{f_{1}(\phi)}{4}F_{(1) MN}F^{MN}- \frac{f_{2}(\phi)}{4}F_{(2)MN}F^{MN} -\frac{1}{2}\partial_{M}\phi \partial^{M}\phi -V(\phi)\right]\,.
\label{actionEF}
\end{eqnarray}
where $F_{(1)MN}$ and $F_{(2)MN}$ are the field strength tensors for two $U(1)$ gauge fields, $\phi$ is the dilaton field,  $f_{1}(\phi)$ and $f_{2}(\phi)$ are the gauge kinetic functions representing the coupling between the two $U(1)$ gauge fields on one hand and the dilaton on the other hand. $V(\phi)$ is the potential of the dilaton field, whose explicit form will depend on the scale function $A(z)$ (see below), and $G_5$ is the Newton constant in five dimensions.  The inspiration for this kind of modeling came from \cite{Arefeva:2018hyo}, albeit that the latter work happened in a different context. Concerning the interpretation of the Abelian gauge fields, we can consider $A_1$ as the dual of a (neutral) flavor current, capable of creating mesons, while $A_2$ is the dual of the electromagnetic current. In principle, the latter can create a different neutral meson. Since we work with $U(1) \times U(1)$, the mesons are charge neutral, so we cannot directly couple electromagnetism to them.   Indeed, we do not have a direct coupling between the 2 gauge fields, so we will never be able to couple e.g.~a magnetic field $B$ to the neutral meson(s). For our current purposes, we will employ the second gauge field just to introduce a (constant) magnetic field $B$, i.e.~we have no interest in the fluctuations of $A_2$. Notice that this $B$ is the 5-dimensional magnetic field, that needs to be suitably rescaled via the AdS length $L$ to get the physical, 4-dimensional, magnetic field $\cal B$. How to do this can be found in \cite{Dudal:2015wfn}. As we are mostly interested in qualitative features in terms of the magnetic field, we will deliberately keep using the 5-dimensional $B$.

One major drawback of the action (\ref{actionEF}) is that it neither explicitly incorporates the dynamics of the chiral condensate nor that it directly couples the chiral condensate to the magnetic field. Consequently, the effects of the magnetic field on the chiral condensate can only be incorporated indirectly from the background metric, see also \cite{Dudal:2015wfn}. Therefore, contrary to real QCD, the backreaction of the chiral condensate on the quark-antiquark free energy and vice versa will be completely ignored in the holographic model (\ref{actionEF}). This is a major disadvantage of all probe-approximated AdS/QCD models, like eq.~(\ref{actionEF}),  where no explicit interplay between the chiral condensate and Polyakov loop exists.  A more accurate and realistic holographic QCD model, that incorporates the backreaction of the chiral field on the spacetime geometry from the start, would---although very interesting---be extremely challenging to construct analytically. We are thus working with a kind of holographic analogue of the quenched QCD approximation known from lattice QCD (no dynamical quarks). As such, one might question how to even couple the magnetic field to the theory if there are no dynamical charge carriers. Here, we follow the pragmatic approach of e.g.~\cite{Critelli:2016cvq,Gursoy:2018ydrbis}, in itself magnetic field-dependent generalizations of the seminal works \cite{Gubser:2008ny,Gubser:2008yx}, and we thus consider our engineered boundary model capable of mimicking some essential QCD features, after which it can be used to describe, without further input, other QCD-ish properties. Notice that Einstein-Maxwell-dilaton models have been used throughout literature as an effective way to describe QCD in presence of electromagnetic background fields, as it is evident from our extensive reference list.

By varying the action (\ref{actionEF}) one can derive the equations of motion for Einstein, Maxwell and dilaton fields. Since, in this work we are mostly interested in a magnetised black brane solution with running dilaton, we consider the following Ans\"atze for the metric $g_{MN}$, field strength tensor $F_{(i)MN}$ and dilaton field $\phi$,
\begin{eqnarray}
& & ds^2=\frac{L^2 S(z)}{z^2}\biggl[-g(z)dt^2 + \frac{dz^2}{g(z)} + dy_{1}^2+ e^{B^2 z^2} \biggl( dy_{2}^2 + dy_{3}^2 \biggr) \biggr]\,, \nonumber \\
& & \phi=\phi(z), \ \ A_{(1) M}=A_{t}(z)\delta_{M}^{t}, \ \  F_{(2)MN}=B dy_{2}\wedge dy_{3}\,.
\label{ansatz}
\end{eqnarray}
where $S(z)$ is the scale factor, $L$ is the AdS length scale and $g(z)$ is the blackening function. $z$ is the usual holographic radial coordinate, and in our coordinate system it runs from $z=0$ (asymptotic boundary) to $z=z_h$ (horizon radius), or to $z=\infty$ for thermal AdS (without horizon). We introduced a background magnetic field $B$ in the $y_1$-direction. Because of this background magnetic field, the system no longer enjoys the $SO(3)$ invariance in boundary spatial coordinates ($y_1$, $y_2$, $y_3$), and we precisely chose the metric Ans\"atze such that as soon as we switch off the magnetic field the $SO(3)$ invariance is recovered.

Using the Ans\"atze of eq.~(\ref{ansatz}) we get four Einstein equations of motion,
\begin{eqnarray}
g''(z) +g'(z) \left(2 B^2
   z+\frac{3 S'(z)}{2 S(z)}-\frac{3}{z}\right) -\frac{z^2 f_{1}(z) A_{t}'(z)^2}{L^2 S(z)} = 0\,.
\label{EOM11}
\end{eqnarray}
\begin{eqnarray}
\frac{B^2 z e^{-2 B^2 z^2} f_{2}(z)}{L^2 S(z)}+2 B^2 g'(z)+g(z)
   \left(4 B^4 z+\frac{3 B^2 S'(z)}{S(z)}-\frac{4 B^2}{z}\right) = 0 \,.
\label{EOM22}
\end{eqnarray}
\begin{eqnarray}
S''(z)-\frac{3 S'(z)^2}{2 S(z)}+\frac{2 S'(z)}{z} + S(z) \left(\frac{4 B^4 z^2}{3}+\frac{4 B^2}{3}+\frac{1}{3} \phi
   '(z)^2\right) = 0 \,.
\label{EOM33}
\end{eqnarray}
\begin{eqnarray}
\frac{g''(z)}{3g(z)} +\frac{S''(z)}{S(z)} +S'(z) \left(\frac{7 B^2 z}{2 S(z)}+\frac{3 g'(z)}{2 g(z)
   S(z)}-\frac{6}{z S(z)}\right) + g'(z) \left(\frac{5 B^2 z}{3
   g(z)}-\frac{3}{z g(z)}\right)  \nonumber \\
     + 2 B^4 z^2+\frac{B^2 z^2 e^{-2 B^2 z^2} f_{2}(z)}{6 L^2 g(z)
   S(z)}-6 B^2+\frac{2 L^2 S(z) V(z)}{3 z^2
   g(z)}+\frac{S'(z)^2}{2 S(z)^2}+\frac{8}{z^2} = 0 \,.
\label{EOM44}
\end{eqnarray}
Similarly we get the following equation of motion for the dilaton field,
\begin{eqnarray}
 \phi ''(z) +\phi '(z) \left(2 B^2
   z+\frac{g'(z)}{g(z)}+\frac{3 S'(z)}{2
   S(z)}-\frac{3}{z}\right) + \frac{z^2 A_{t}'(z)^2}{2 L^2 g(z)
   S(z)}\frac{\partial f_{1}(\phi)}{\partial \phi}  \nonumber \\
   -\frac{B^2 z^2 e^{-2 B^2 z^2}}{2 L^2 g(z) S(z)} \frac{\partial f_{2}(\phi)}{\partial \phi} -\frac{L^2 S(z)}{z^2 g(z)} \frac{\partial V(\phi)}{\partial \phi} =0\,.
\label{dilatonEOM}
\end{eqnarray}
and the equation of motion for the first gauge field,
\begin{eqnarray}
A_{t}''(z)+ A_{t}'(z) \left(2 B^2
   z+\frac{f_{1}'(z)}{f_{1}(z)}+\frac{S'(z)}{2
   S(z)}-\frac{1}{z}\right) =0\,.
\label{MaxwellAtEOM}
\end{eqnarray}
One can explicitly check that the equation of motion for the second Maxwell field is trivially satisfied and hence it will not give any additional equation. Therefore, we have in total six equations of motion. However, only five of them independent. Below we will choose the dilaton equation (\ref{dilatonEOM}) as a constrained equation and consider the rest of the equations as independent. In order to solve the latter, we impose the following boundary conditions,
\begin{eqnarray}
&& g(0)=1 \ \ \text{and} \ \ g(z_h)=0, \nonumber \\
&& A_{t}(0)= \mu \ \ \text{and} \ \  A_{t}(z_h)=0, \nonumber \\
&& S(0) = 1, \nonumber \\
&& \phi(0)=0\,.
\label{boundaryconditions}
\end{eqnarray}
where $\mu$ is the chemical potential of the boundary theory which is related to the near boundary expansion of the zeroth component of the first gauge field and, as mentioned before, $z_h$ is the location of the black hole horizon. Apart from these boundary conditions, we will also assume that the dilaton field $\phi$ remains real everywhere in the bulk. As we will see later, this condition will severely restrict our analytic solution for a finite magnetic field.\\

In order to solve eqs.~(\ref{EOM11}), (\ref{EOM22}), (\ref{EOM33}), (\ref{EOM44}) and (\ref{MaxwellAtEOM}) simultaneously, we adopt the following strategy.
\begin{enumerate}
\item We first solve eq.~(\ref{MaxwellAtEOM}) and obtain the solution for $A_t(z)$ in terms of $f_1(z)$ and $S(z)$.

\item Using $A_t(z)$, we then solve eq.~(\ref{EOM11}) and find the solution for $g(z)$ in terms of $f_1(z)$ and $S(z)$.

\item Using $g(z)$, we then solve eq.~(\ref{EOM22}) to obtain $f_2(z)$.

\item Next, we solve eq.~(\ref{EOM33}) and find $\phi'(z)$ in terms of $S(z)$.

\item Finally, we solve eq.~(\ref{EOM44}) and obtain the dilaton potential in terms of $S(z)$ and $g(z)$.

\end{enumerate}
Applying the above strategy and solving eq.~(\ref{MaxwellAtEOM}), we get the following solution for $A_t$
\begin{eqnarray}
A_{t} (z) = K_{1} \int_0^z \, d\xi \frac{ \xi e^{-B^2 \xi^2}}{f_{1}(\xi)
   \sqrt{S(\xi)}} +K_2 \,.
\end{eqnarray}
Applying the boundary condition (eq.~(\ref{boundaryconditions})), we get
\begin{eqnarray}
K_2 = \mu,  \ \ \ \ \ K_1 = -\frac{\mu}{ \int_0^{z_h} \, d\xi \frac{ \xi e^{-B^2 \xi^2}}{f_{1}(\xi)
   \sqrt{S(\xi)}}}  \,.
\end{eqnarray}
and the solution for $A_{t}$ then becomes,
\begin{eqnarray}
A_{t} (z) = \mu \biggl[ 1 - \frac{ \int_0^z \, d\xi \frac{ \xi e^{-B^2 \xi^2}}{f_{1}(\xi)
   \sqrt{S(\xi)}}}{ \int_0^{z_h} \, d\xi \frac{ \xi e^{-B^2 \xi^2}}{f_{1}(\xi)
   \sqrt{S(\xi)}}}  \biggr] = \tilde{\mu} \int_z^{z_h} \, d\xi \frac{ \xi e^{-B^2 \xi^2}}{f_{1}(\xi)
   \sqrt{S(\xi)}} \,.
 \label{Atsol}
\end{eqnarray}
Substituting eq.~(\ref{Atsol}) into eq.~(\ref{EOM11}), we get the following solution for $g(z)$,
\begin{eqnarray}
g(z) = \int_0^z \, d\xi \frac{ \xi^3 e^{-B^2 \xi^2}}{\sqrt{S^{3}(\xi)}} \biggl[ K_{3} + \frac{\tilde{\mu}^2}{L^2} \int_0^\xi \, d\tilde{\xi} \frac{ \tilde{\xi} e^{-B^2 \tilde{\xi}^2}}{f_{1}(\tilde{\xi})\sqrt{S(\tilde{\xi})}} \biggr] +    K_4 \,.
\label{gsol}
\end{eqnarray}
The constants $K_3$ and $K_4$ can be fixed from eq.~(\ref{boundaryconditions}) and we get
\begin{eqnarray}
K_4 = 1,  \ \ \ \ \ K_3 = \frac{-1}{ \int_0^{z_h} \, d\xi \frac{ \xi^3 e^{-B^2 \xi^2}}{\sqrt{S^{3}(\xi)}}} \biggl[1+\frac{\tilde{\mu}^2}{L^2}
\int_0^{z_h} \, d\xi \frac{ \xi^3 e^{-B^2 \xi^2}}{\sqrt{S^{3}(\xi)}} \biggl( \int_0^\xi \, d\tilde{\xi} \frac{\tilde{\xi} e^{-B^2 \tilde{\xi}^2}}{f_{1}(\tilde{\xi})\sqrt{S(\tilde{\xi})}}  \biggr) \biggr]  \,.
\end{eqnarray}
The coupling function can be obtained from eq.~(\ref{EOM22}),
\begin{eqnarray}
f_2(z) = - \frac{ e^{2 B^2 z^2} L^2 S(z)}{z}\biggl[ g(z) \left(4 B^2 z+\frac{3
   S'(z)}{S(z)}-\frac{4}{z}\right)+2 g'(z) \biggr] \,.
\label{f2sol}
\end{eqnarray}
Similarly, the dilaton field can be obtained from eq.~(\ref{EOM33})
\begin{eqnarray}
\phi'(z) &=& \frac{\sqrt{-8 B^4 z^3 S(z)^2-8 B^2 z S(z)^2-6 z S(z) S''(z)+9 z S'(z)^2-12
   S(z) S'(z)}}{\sqrt{2 z} S(z)}, \nonumber \\
\phi(z) &=&  \int \, dz  \frac{\sqrt{-8 B^4 z^3 S(z)^2-8 B^2 z S(z)^2-6 z S(z) S''(z)+9 z S'(z)^2-12
   S(z) S'(z)}}{\sqrt{2 z} S(z)} + K_{5}
\label{phisol}
\end{eqnarray}
where the constant $K_{5}$ will be fixed demanding that\footnote{This simple choice assures we asymptote back to AdS$_5$ near the UV QCD boundary $z=0$.} $\phi |_{z=0}\rightarrow 0$. And finally, eq.~(\ref{EOM44}) allows us to find the potential,
\begin{eqnarray}
V(z) = \frac{g(z)}{L^2} \left(-\frac{9 B^2 z^3 S'(z)}{2 S(z)^2}+\frac{10
   B^2 z^2}{S(z)}-\frac{3 z^2 S'(z)^2}{S(z)^3}+\frac{12 z
   S'(z)}{S(z)^2}+\frac{z^2 \phi '(z)^2}{2 S(z)}-\frac{12}{S(z)}\right)  \nonumber \\
   -\frac{z^4 f_{1}(z)A_{t}'(z)^2}{2 L^4 S(z)^2}+ \frac{g'(z)}{L^2}
   \left(-\frac{B^2 z^3}{S(z)}-\frac{3 z^2 S'(z)}{2 S(z)^2}+\frac{3
   z}{S(z)}\right)  \,.
\label{Vsol}
\end{eqnarray}
It is clear from the above equations that a complete analytic solution to the Einstein-Maxwell-dilaton system with a background magnetic field can be obtained in terms of two arbitrary functions, i.e.~the scale function $S(z)$ and the gauge coupling $f_1(z)$. Different forms of $S(z)$ and $f_1(z)$ will give different physically allowed solutions. Indeed, it can be explicitly verified that the Einstein, Maxwell and dilaton equations are satisfied for any form of $S(z)$ and $f_1(z)$. Thus we have found a family of analytic solutions for the gravity system of eq.~(\ref{actionEF}). Since our aim here is to model real QCD properties holographically, we will fix these two arbitrary functions by taking inputs from real QCD. For example, by comparing the holographic results for the deconfinement transition temperature and meson mass spectrum with lattice QCD, we can fix/constrain the forms of $S(z)$ and $f_1(z)$.

The form of the gauge coupling function $f_1$ can be constrained by studying the vector meson mass spectrum. In particular, by taking the following simple form of $f_1$,
\begin{eqnarray}
f_1(z) = \frac{e^{-c z^2 -B^2 z^2}}{\sqrt{S(z)}} \,.
\label{f1sol}
\end{eqnarray}
the vector meson spectra can be shown to lie on linear Regge trajectories for $B=0$. In particular, the mass squared of the vector mesons satisfies $m_{n}^{2} = 4cn$. Moreover, the parameter $c$ can also be fixed by matching with lowest lying heavy meson states $J/\Psi$ and $\Psi'$, and by doing that we get $c=1.16~\text{GeV}^{2}$, see \cite{Yang:2015aia,Dudal:2017max} for more details.

Substituting eq. (\ref{f1sol}) into eqs. (\ref{gsol}), (\ref{f2sol}) and (\ref{phisol}), and using $S(z)=e^{2 A(z)}$, we get the following solutions,
\begin{eqnarray}
g(z) &=& 1 +  \int_0^z \, d\xi \ \xi^3 e^{-B^2 \xi^2 -3A(\xi) } \biggl[ K_{3} + \frac{\tilde{\mu}^2}{2 c L^2} e^{c \xi^2}  \biggr], \nonumber \\
&&\text{with} \ \ \ K_3 = - \frac{ \biggl[1+\frac{\tilde{\mu}^2}{2 c L^2} \int_0^{z_h} \, d\xi \  \xi^3 e^{-B^2 \xi^2-3A(\xi)+c \xi^2}
 \biggr]}{\int_0^{z_h} \, d\xi \ \xi^3 e^{-B^2 \xi^2-3A(\xi)}}  \,.
\label{gsol1}\\
f_2(z) &=& - \frac{ L^2 e^{2 B^2 z^2 +2A(z)} }{z}\biggl[ g(z) \left(4 B^2 z+ 6 A'(z)-\frac{4}{z}\right)+2 g'(z) \biggr] \,.
\label{f2sol1}\\
\phi(z) &=& \int \, dz \sqrt{-\frac{2}{z} \left(3 z A''(z)-3 z A'(z)^2+6
   A'(z)+2 B^4 z^3+2 B^2 z\right)} +K_5  \,.
\label{phisol1}
\end{eqnarray}
Notice that $\sqrt{S(z)}$, appearing in eq.~\eqref{f1sol}, is then well-defined. Also, one can substitute eq. (\ref{f1sol}) into eqs. (\ref{Atsol}) and (\ref{Vsol}) to obtain the explicit solutions for $A_{t}(z)$ and $V(z)$. Therefore, in  eqs.~ (\ref{Atsol}), (\ref{gsol}), (\ref{f2sol}), (\ref{phisol}) and (\ref{Vsol}) a complete solution for the Einstein-Maxwell-dilaton gravity system is obtained in terms of a single scale function $A(z)$.

Let us also note the expressions of black hole temperature and entropy, which will be useful later on in the investigation of the black hole thermodynamics
\begin{eqnarray}
T &=& - \frac{z_h^3 \ e^{-3 A\left(z_h\right)-B^2 z_h^2}}{4 \pi} \biggl[K_3 + \frac{\tilde{\mu}^2}{2 c L^2} e^{c z_{h}^2} \biggr] ,  \  \  \ S = \frac{e^{B^2 z_{h}^{2}+3A(z_h)}}{4 z_{h}^3}.
\label{phicase11}
\end{eqnarray}

Before we close this section, it is important to emphasize again that eqs.~(\ref{Vsol})-(\ref{phisol1}) are a solution of the action (\ref{actionEF}) for any scale factor $A(z)$. We therefore have an infinite family of analytic black hole solutions for the gravity system of eq.~(\ref{actionEF}). These different solutions however correspond to
different dilaton potentials (and therefore to different actions), as different forms of $A(z)$ will give different potentials $V(z)$. However, once the form of $A(z)$ is fixed then the form of $V(z)$ also is, and in return  eqs.~(\ref{Vsol})-(\ref{phisol1}) correspond to a self-consistent solution to a particular action with predetermined $A(z)$ and $V(z)$.

One might also worry that the dependence of $V$ on the parameters $z_h$, $\mu$ and $B$ is troublesome as it indicates that different values of these parameters correspond to a different action, and therefore, to a different gravity model altogether. We like to emphasise here that $V$ does not depend on these parameters explicitly at the level of the action or equations of motion. These parameters appear only when the boundary conditions in eq.~(\ref{boundaryconditions}) are imposed. Indeed, notice that there is a second independent solution to our EMD equations of motion, which corresponds to thermal-AdS. For the thermal-AdS, we have $g(z)=1$. Then it is easy to infer from eq.~(\ref{EOM44}) that $V$ is independent of $z_h$ and $\mu$ (the $B$ dependence appears because of the metric and gauge field ans\"atze). If the potential were dependent on $z_h$ and $\mu$ from the beginning in the action itself, then the potential in the thermal-AdS background would have depended on $z_h$ and $\mu$ as well, which is certainly not the case in our model as we have alluded to  above. Therefore, one should interpret the dependence of the potential $V$ on $z_h$ and $\mu$ as an on-shell dependence, and not as an off-shell one. In any case, we have numerically checked that the on-shell $V$ depends only very mildly on $z_h$, $\mu$ and $B$. In particular, the potential profiles for different $z_h$, $\mu$ and $B$ values are almost indistinguishable from each other in the region away from the horizon whereas they are separable only mildly in the near horizon region. Illustrative figures are included in Appendix A.

\section{Results}
Following \cite{Dudal:2017max}, we will depart from the $B=0$ Ansatz\footnote{In ongoing work, a more general form factor will be employed so that next to a confining linear potential, also asymptotic freedom can be built in, see e.g.~\cite{Megias:2010ku,Gursoy:2007er}. In any case, our main results concerning the anisotropic string tension will remain the same even after employing a more sophisticated form factor which will guarantee asymptotic freedom in UV.}
\begin{eqnarray}
A(z) = - a z^2\,.
\label{Acase1}
\end{eqnarray}
Let us first note the expression of $\phi(z)$
\begin{eqnarray}
\phi(z) = \frac{\left(9 a-B^2\right) \log \left(\sqrt{6
   a^2-B^4} \sqrt{6 a^2 z^2+9 a - B^4 z^2 -B^2}+6 a^2 z - B^4 z
   \right)}{\sqrt{6 a^2-B^4}} \nonumber \\
   +z \sqrt{6 a^2 z^2+9 a - B^2 \left(B^2
   z^2+1\right)} - \frac{\left(9 a-B^2\right) \log \left(\sqrt{9 a-B^2} \sqrt{6
   a^2-B^4}\right)}{\sqrt{6 a^2-B^4}} \,.
\label{phicase11}
\end{eqnarray}
Similar expressions can be found for $A_t(z)$, $g(z)$, $f_2(z)$ and $V(z)$ as well. However, these expressions are too lengthy to reproduce here and also not particularly illuminating, we therefore just mentioned the functionality of $\phi(z)$ since it gives the stability criteria of our solution. Indeed, from eq.~(\ref{phicase11}), we learn that the dilaton field is real-valued only when $B^4\leq 6a^2$. This condition severely restricts the validity of our gravity solution and below we will work with only those values of $a$ and $B$ for which this condition is satisfied. Moreover, it is also interesting to note that the Gubser criterion \cite{Gubser:2000nd}---the scalar potential must be bounded from above, $V(z)\leq V(0)$, for a physically acceptable holographic solution---is always satisfied under the same condition $B^4\leq6a^2$.  In particular, the potential is almost constant having value $-12/L^2$ near the asymptotic boundary and then decreases in the deep IR. This, therefore, gives a strong self-consistency check on our constructed solution.

\subsection{Black hole thermodynamics and confinement-deconfinement phase transition}
\begin{figure}[h!]
\begin{minipage}[b]{0.5\linewidth}
\centering
\includegraphics[width=2.8in,height=2.3in]{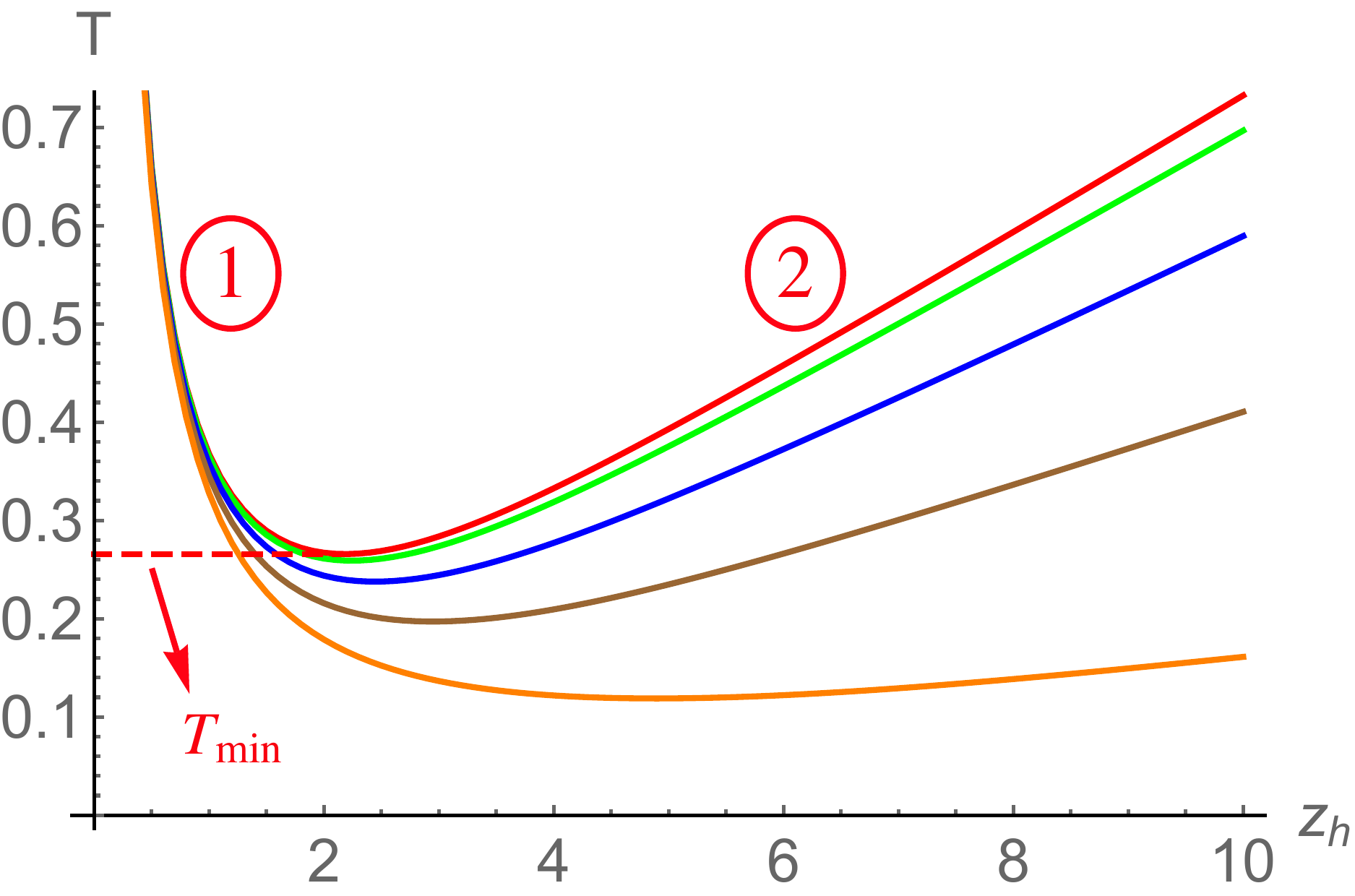}
\caption{ \small Temperature $T$ as a function of horizon radius $z_h$ for various values of the magnetic field $B$ and $\mu=0$.  Here red, green, blue, brown and orange curves correspond to $B=0$, $0.15$, $0.30$, $0.45$ and $0.6$ respectively. In units \text{GeV}.}
\label{zhvsTvsBMu0case1}
\end{minipage}
\hspace{0.4cm}
\begin{minipage}[b]{0.5\linewidth}
\centering
\includegraphics[width=2.8in,height=2.3in]{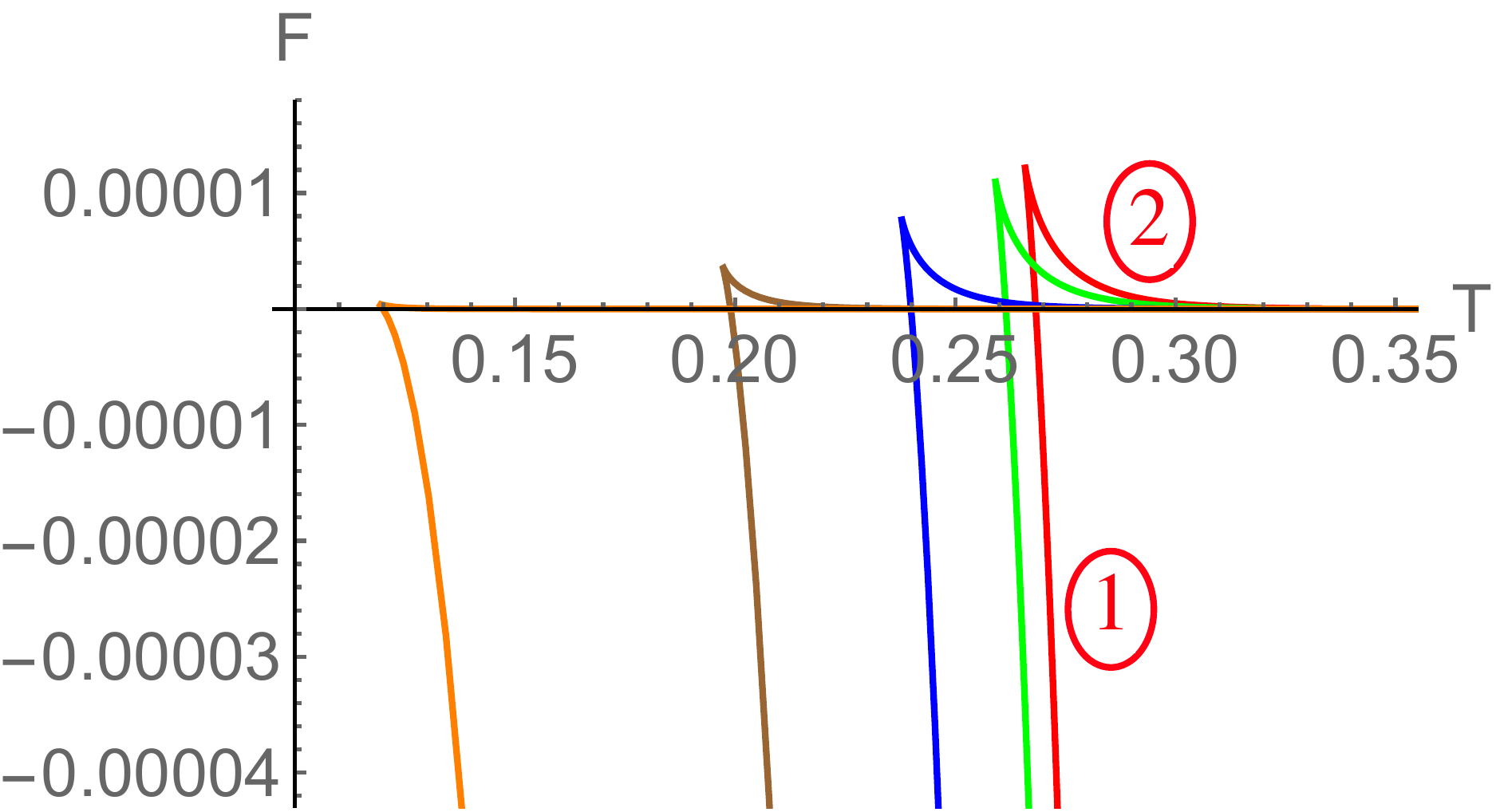}
\caption{\small Free energy $F$ as a function of temperature $T$ for various values of the magnetic field $B$ and $\mu=0$. Here red, green, blue, brown and orange curves correspond to $B=0$, $0.15$, $0.30$, $0.45$ and $0.6$ respectively. In units \text{GeV}.}
\label{TvsFvsBMu0case1}
\end{minipage}
\end{figure}

The thermodynamics of the gravity solution with the scale factor of eq.~(\ref{Acase1}) is shown in Figures~\ref{zhvsTvsBMu0case1} and  \ref{TvsFvsBMu0case1}. In Figure \ref{zhvsTvsBMu0case1}, the variation of Hawking temperature $T$ with respect to the horizon radius $z_h$ for various values of the magnetic field $B$ is shown. We find that there exists a minimum temperature $T_{min}$ below which no black hole solution exist. However, for $T>T_{min}$, there are two black hole solutions, a large and a small one, which are marked by $\circled{1}$ and $\circled{2}$ respectively. The small black hole phase for which $T$ increases with $z_h$ is unstable whereas the large black hole phase for which $T$ decreases with $z_h$ is stable. The unstable-stable nature of the small-large black hole phases can be seen from the free energy behaviour shown in Figure~\ref{TvsFvsBMu0case1}. Here, we have normalised the free energy of the black hole with respect to the thermal AdS case, $z_h\to\infty$. We see that the free energy of the small black hole phase is always larger than the large black hole and thermal AdS phases, indicating the unstable nature of this small black hole phase. Importantly, upon varying the Hawking temperature, a phase transition from the large black hole phase to thermal AdS phase takes place at a critical temperature $T_{crit}$.  This is the famous black hole-thermal AdS Hawking-Page phase transition \cite{Hawking}.

\begin{figure}[h!]
\begin{minipage}[b]{0.5\linewidth}
\centering
\includegraphics[width=2.8in,height=2.3in]{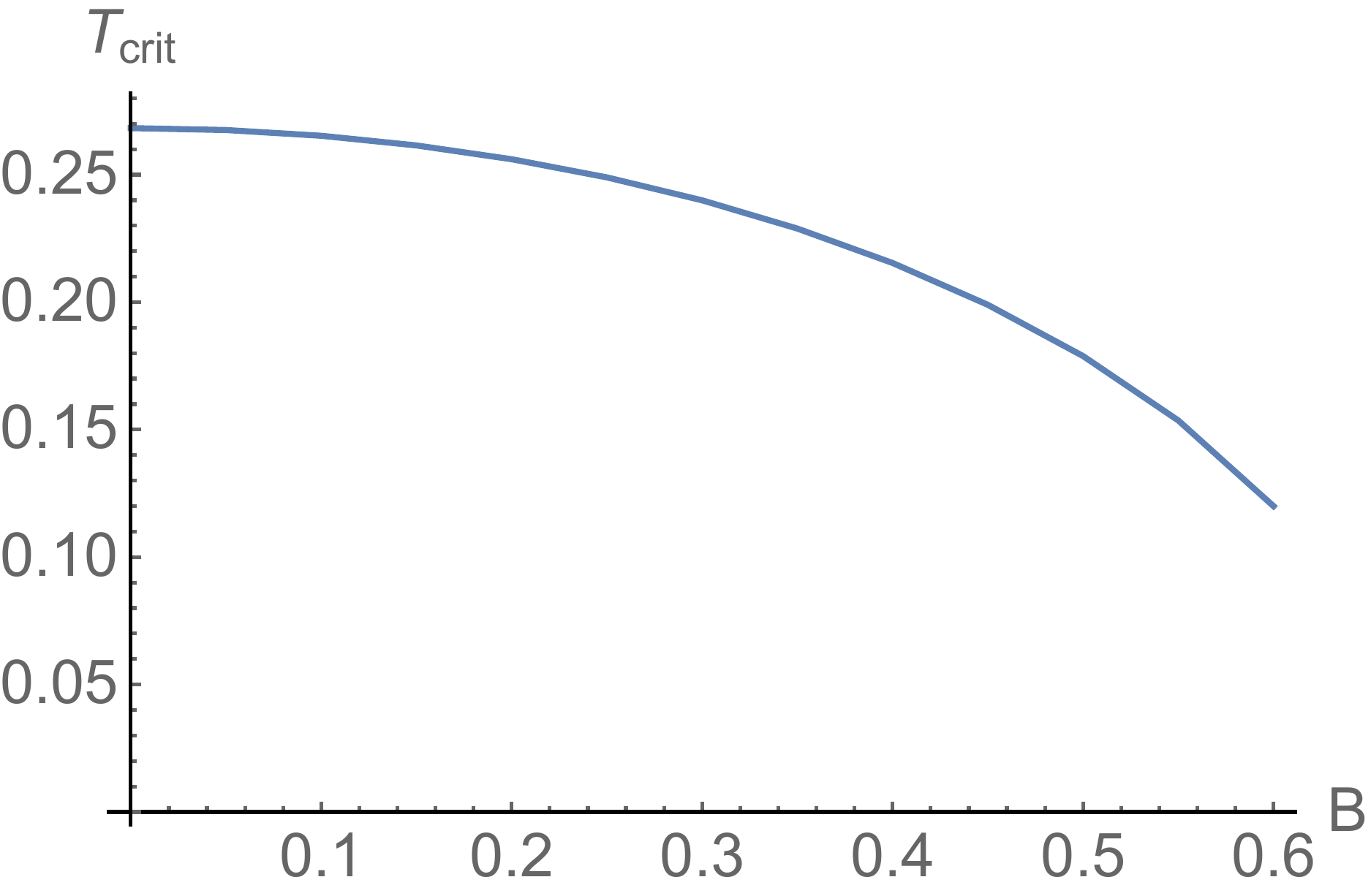}
\caption{ \small The variation of thermal AdS--black hole phase transition critical temperature $T_{crit}$ with magnetic field $B$. Here $\mu=0$ is considered. In units \text{GeV}.}
\label{TcritvsBMu0case1}
\end{minipage}
\hspace{0.4cm}
\begin{minipage}[b]{0.5\linewidth}
\centering
\includegraphics[width=2.8in,height=2.3in]{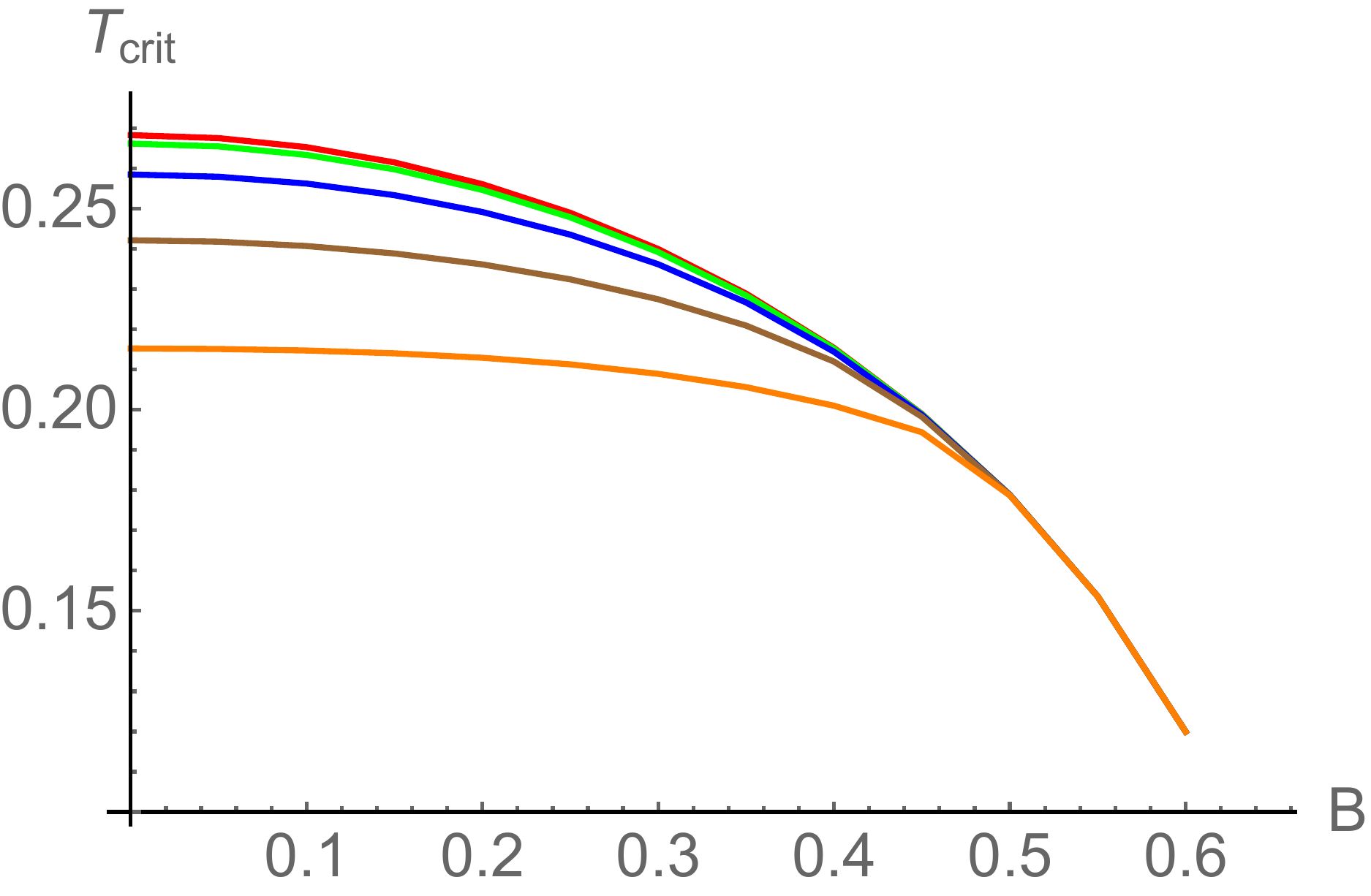}
\caption{\small The variation of thermal AdS--black hole phase transition critical temperature $T_{crit}$ with magnetic field $B$ for various values of chemical potential $\mu$. Here red, green, blue, brown and orange curves corresponds to $\mu=0.0$, $0.3$, $0.6$, $0.9$ and $1.2$ respectively. In units \text{GeV}.}
\label{TcritvsBvsMucase1}
\end{minipage}
\end{figure}

Interestingly, the above thermodynamic behaviour occurs for small but finite values of the magnetic field as well. For finite magnetic field, we again find the unstable small-stable large black hole phases, with thermal AdS dominating the physics at small temperatures. Importantly, the Hawking-Page thermal AdS--black hole phase transition persists even for finite values of the magnetic field. The main difference appears in the magnitude of the critical temperature $T_{crit}$. In particular, $T_{crit}$ decreases for higher values of magnetic field. The dependence of $T_{crit}$ on $B$ is shown in Figure \ref{TcritvsBMu0case1}, which is also one of the main results of this paper. Since these thermal AdS and black hole phases in the usual language of gauge-gravity duality are dual to the confinement and deconfinement phases in the dual boundary theory, accordingly, our result in Figure \ref{TcritvsBMu0case1} predicts inverse magnetic catalysis for the dual confinement-deconfinement transition. Our result in Figure \ref{TcritvsBMu0case1}, therefore, provides a major improvement on several soft and hard wall models of holographic QCD, which did already suggest inverse magnetic catalysis in the deconfinement sector, however based on a ad hoc choice of the dilaton. Here, we have explicitly included the backreaction of the dilaton field in a not overly complicated way.

One can naively expect by looking at Figures \ref{zhvsTvsBMu0case1} and \ref{TvsFvsBMu0case1} that the above scenario for the thermal AdS--black hole phase transition might change for large magnetic fields. However, we need to be careful here. As mentioned before, the dilaton field, and therefore our gravity solution,  only makes sense when the condition $B^4<B_{c}^{4}=6a^2$ is satisfied. Since we took $a=0.15~\text{GeV}^{2}$ for a decent match with the lattice QCD deconfinement temperature at $B=0$ \cite{Dudal:2017max}, our gravity solution is trustworthy only for $B_c \simeq 0.61~\text{GeV}$. Then we find that for all $B\leq B_c$, the thermal AdS--black hole phase transition occurs. We have also explicitly checked that it persists in terms of a varying $a$. Moreover, the critical temperature $T_{crit}$ decreases with magnetic field even for these different values of $a$, indicating the inverse magnetic catalysis again, in line with independent lattice QCD predictions.

We now move on to discuss the thermodynamic behaviour in the presence of chemical potential $\mu$. We find similar results (as discussed above) with finite $\mu$ as well, and therefore, we can be very brief here. Our results are summarized in Figure \ref{TcritvsBvsMucase1}. One of the main outcomes here is that our holographic model continues to exhibit inverse magnetic catalysis behaviour for non-zero values of $\mu$ as well. Unfortunately, we do not have lattice results for (inverse) magnetic catalysis at finite $\mu$ yet, as lattice simulations usually suffer from the sign problem with $\mu$. Therefore, the results in Figure \ref{TcritvsBvsMucase1} can be considered as a genuine prediction of our holographic model of eq.~\eqref{actionEF}.

Moreover, we also find, like many other holographic QCD models, a decreasing pattern for the critical temperature with chemical potential.

\subsection{The anisotropic QCD string tension}
In order to study the QCD string tension, our approach is to consider the free energy $\cal F$ of a $q,\bar q$ pair via the dual of the gauge invariant quantity from which the $q,\bar q$ interaction energy can be extracted \cite{Bali:2000gf}, i.e.~via the holographic realization of the Wilson loop \cite{Maldacena:1998im,Brandhuber:1998bs,CasalderreySolana:2011us,Andreev:2006ct}. In particular,
the gauge-gravity duality provides a correspondence between the $\cal F$ of the $q, \bar q$ pair with separation distance $\ell$ that evolves over a large time $T$ and the Nambu-Goto (NG) on-shell action \footnote{In principle, because of the non-trivial dilaton profile, there can be an additional term in the NG string world sheet action that describes the coupling between the dilaton field and two-dimensional Ricci scalar of the world sheet. However, in the large 't Hooft limit $\lambda\rightarrow\infty$---with which one is implicitly always working  in the gauge-gravity correspondence---this term will be negligible being an $\mathcal{O}(\alpha')\sim\frac{1}{\sqrt{\lambda}}$ contribution, and therefore it is always omitted in any kind of applied gauge/gravity computation.}. This action describes the physics of the open string that evolves in time and sweeps out a two dimensional world-sheet which is bounded on the AdS boundary by the rectangular Wilson loop $\ell\times T$. So, we have
\begin{equation}
{\cal{F}} (\ell ,T)= T S^{on-shell}_{NG}(\ell ,T)~,\,
\label{nambugoto}
\end{equation}
where
\begin{eqnarray}
S_{NG}=\frac{1}{2 \pi \ell^2_{s}}\int d\tau d\sigma \sqrt{-\det ~G_{s}}~.
\label{nambugoto2}
\end{eqnarray}
Here, $T_{s}=\frac{1}{2 \pi \ell^2_{s}}$ is the open string tension, the coordinates $(\tau, \sigma)$ are used to parameterize the two-dimensional world-sheet and $(G_{s})_{\alpha \beta}=(g_{s})_{MN} \partial_{\alpha} X^M \partial_{\beta} X^N$, where $X^M(\tau, \sigma)$ indicates the embedding of the open string in the gravity background, $g_{s}$ is the background metric in the string frame\footnote{Hereafter, we use the subscript ``$s$'' to indicate that we are working in the string frame.}, as appropriate to extract the string ($q,\bar q$) free energy \cite{Critelli:2016cvq,Gursoy:2010fj}. $G_{s}$ is the induced metric on the two-dimensional world-sheet.

The above metric solution (\ref{ansatz}) is in the Einstein frame, and the standard method to pass to the string frame can be obtained from the dilaton transformation \cite{Critelli:2016cvq,Dudal:2017max}, i.e.~$(g_{s})_{MN}=e^{\sqrt{\frac{2}{3}}\phi} g_{MN}$. So, the metric solution (\ref{ansatz}) in the string frame is,
\begin{eqnarray}
 ds^2_{s}=\frac{L^2 e^{2A_{s}(z)}}{z^2}\biggl[-g(z)dt^2 + \frac{dz^2}{g(z)} + dy_{1}^2+ e^{B^2 z^2} \biggl( dy_{2}^2 + dy_{3}^2 \biggr) \biggr]\,,
\label{metric5}
\end{eqnarray}
where $A_{s}(z)=A(z)+\sqrt{\frac{1}{6}}~\phi(z)$, with $A(z)$ and $\phi(z)$ as given in eqs. (\ref{Acase1}) and (\ref{phicase11}), respectively.
In the following, we consider two cases to investigate the effect of magnetic field on the QCD string tension, the parallel case, i.e.~when the $q,\bar q$ pair is oriented parallel to the magnetic field, and then also a perpendicular orientation.

\subsubsection{Parallel case}
In this case, to parameterize the string world-sheet, we use the static gauge, i.e.~$\tau= t$ and $\sigma= y_{1}$. So, one can obtain both connected and disconnected  solutions that minimize the Nambu-Goto action from eqs. (\ref{nambugoto2}) and (\ref{metric5}). The connected solution is a $\cup$-shape open string configuration with endpoints as the $q,\bar q$ pair, so that
\begin{eqnarray}
{\cal {F}}^{\|}_{con}= \frac{L^2}{\pi \ell ^{2}_{s}} \int_{\epsilon}^{z^{\|}_*} dz \frac{z^{\| 2}_*}{z^2}\frac{\sqrt{g(z)}~e^{2A_{s}(z)-2A_{s}(z^{\|}_*)}}{\sqrt{g(z) z^{\| 4}_*  e^{-4A_{s}(z^{\|}_*)} - g(z^{\|}_*) z^4 e^{-4A_{s}(z)}}}~,
\label{paraconn}
\end{eqnarray}
where $z^{\|}_*$ is the turning point of  the $\cup$-shaped open string that stretches from the UV boundary at $z=0$ into the bulk at $z=z^{\|}_*$, while $\epsilon$ is the regulator on the gravity side that corresponds to the UV cut-off on the gauge theory side. The relation between the $q, \bar q$ separation $\ell^{\|}$ and $z^{\|}_*$ reads
\begin{eqnarray}
\ell^{\|} =2 \int_{\epsilon}^{z^{\|}_*} dz \sqrt{\frac{g(z^{\|}_*)}{g(z)}}\frac{z^2~e^{-2A_{s}(z)}}{\sqrt{g(z) z^{\| 4}_*  e^{-4A_{s}(z^{\|}_*)} - g(z^{\|}_*) z^4 e^{-4A_{s}(z)}}}~.
\label{ellconn}
\end{eqnarray}
For the disconnected solution, the free energy of the $q,\bar{q}$ pair becomes
\begin{eqnarray}
{\cal {F}}^{\|}_{discon}= \frac{L^2}{\pi \ell ^{2}_{s}} \int_{\epsilon}^{z_h} dz \frac{e^{2A_{s}(z)}}{z^2}
\label{paraconn}
\end{eqnarray}
where in principle, $z_h$ is the horizon of the black hole. However, as we want to study the effect of magnetic field on the QCD string tension in the confined phase, we send $z_h \to\infty$ to work in the thermal AdS background. In the rest of the section we will work in the thermal AdS background for which $g(z)=1$. As usual, both solutions ${\cal F}^{\|}_{con}$ and ${\cal F}^{\|}_{discon}$ are UV divergent when $\epsilon\rightarrow 0$. In this work, we use the subtraction procedure that was mentioned in \cite{Ewerz:2016zsx}: we minimally\footnote{In particular, we subtracted $\frac{2}{\epsilon}-\sqrt{\frac{32}{3}(9a-B^2)}\log{\epsilon}$ from ${\cal {F}}^{\|}_{con}$ and ${\cal {F}}^{\|}_{discon}$ to get the renormalized $q, \bar q$ free energy. } remove the contribution of pole parts to obtain the renormalized results.

\begin{figure}[h!]
\begin{minipage}[b]{0.5\linewidth}
\centering
\includegraphics[width=2.7in,height=1.9in]{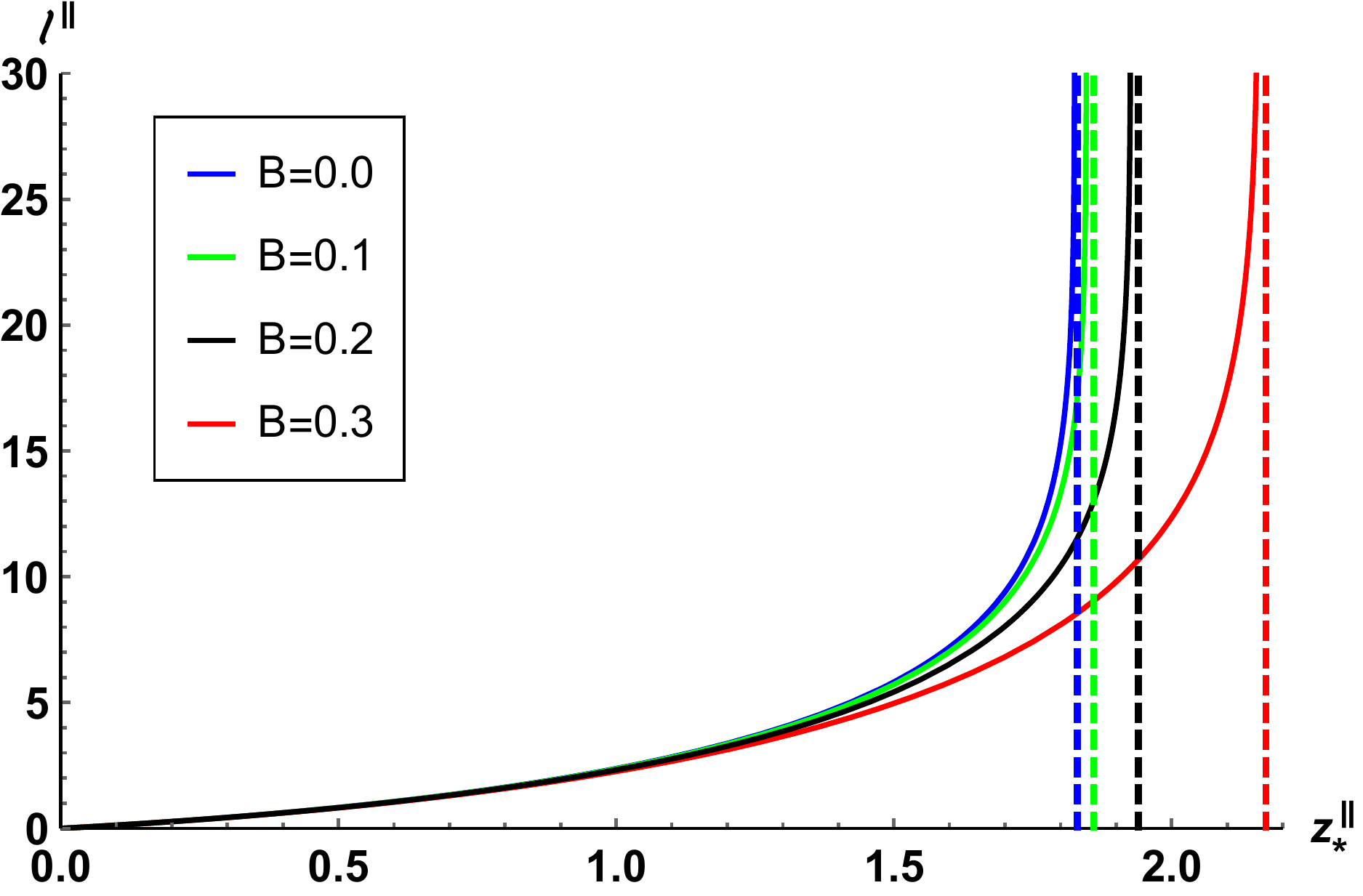}
\caption{ \small $\ell^{\|}$ as a function of $z^{\|}_*$ in the thermal AdS background for different (small) magnetic fields and $\mu=0$. In units \text{GeV}.}
\label{ellparalzstar}
\end{minipage}
\hspace{0.4cm}
\begin{minipage}[b]{0.5\linewidth}
\centering
\includegraphics[width=2.7in,height=1.9in]{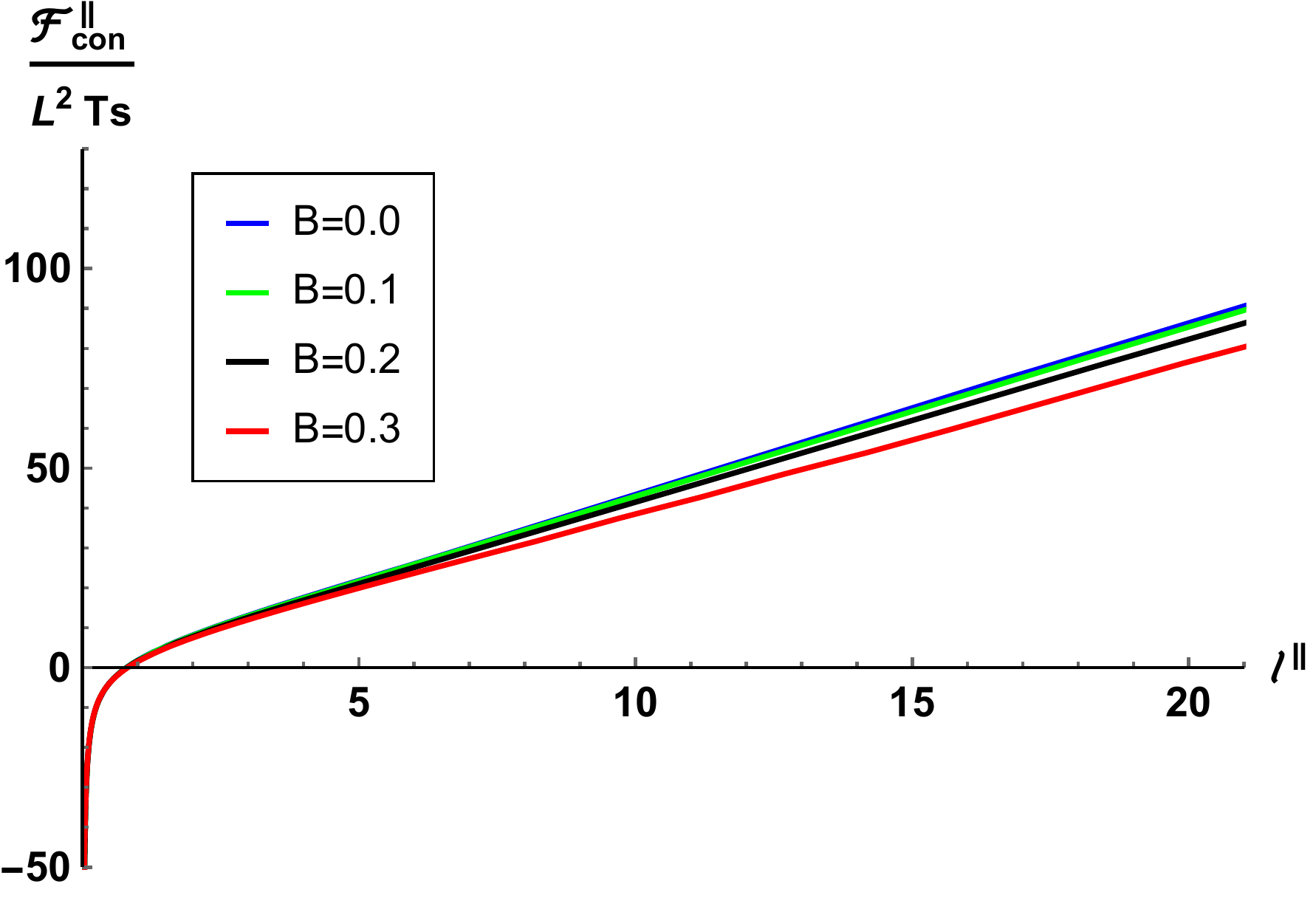}
\caption{ \small  ${\cal {F}}^{\|}_{con}$ as a function of $\ell^{\|}$ in the thermal AdS background for different  (small) magnetic fields and $\mu=0$. In units \text{GeV}.}
\label{fconn1}
\end{minipage}
\end{figure}

Let us first focus on $\ell^{\|}$ vs. $z^{\|}_*$ behaviour for different magnetic fields and $\mu=0$, as shown in Figure \ref{ellparalzstar}. We see that, when we increase $\ell^{\|}$, a dynamical ``imaginary wall'' appears in the bulk spacetime beyond which the connected string world sheet does not propagate. This ``imaginary wall'' appears when the square root in the integrand of $\ell$ can become negative, which is possible when the scale factor of the metric in the string frame \cite{Gursoy:2007er} develops a minimum in which case we will encounter an ``imaginary wall'' that moves with different values of the magnetic field.  A similar type of ``imaginary wall'' has been reported before as well \cite{Dudal:2017max,Yang:2015aia,Arefeva:2018hyo}. This means that the original hard or soft wall of models like \cite{Erlich:2005qh,Karch:2006pv} gets replaced by a dynamical kind of wall giving similar features. We can thus increase the $q, \bar{q}$ separation $\ell$ in such a way that the $q$ and $\bar{q}$ that are connected by the open string, remains bound and thus cannot dissociate. Accordingly, we may then state that the $q, \bar{q}$ are connected to each other and form a confined state on the gauge theory side. In addition, the location of the ``imaginary wall'' shifts to higher values of $z$ by increasing the magnetic field, corresponding to a deeper penetration into the bulk.

Let us now consider the free energy of the $q, \bar{q}$ pair in the connected configuration ${\cal {F}}^{\|}_{con}$ as a function of $\ell^{\|}$ for different magnetic fields in the thermal AdS background, shown in Figure \ref{fconn1}. We find that each of them can be fitted with a Cornell-type potential \cite{Eichten:1974af,Eichten:1978tg},
\begin{eqnarray}
\frac{{\cal {F}}^{\|}_{con}}{L^2 T_{s}}= -\frac{\kappa^\|}{\ell^{\|}}+ {\sigma}_{s}^{\|}{\ell}^{\|} + C^\|
\label{cornell}
\end{eqnarray}
where $\kappa^\|$ is a Coulomb strength parameter, $\sigma_{s}^{\|}$ is the QCD string tension and $C^\|$ is a constant shift in the potential. To be more precise, all these QCD-related quantities are to be rescaled with $T_s L^2$. For small values of $\ell^{\|}$, the Coulomb potential, $-\frac{\kappa^\|}{\ell^{\|}}$ dominates and for larger values of $\ell^{\|}$, the linear potential part $\sigma_{s}^{\|}\ell^{\|}$ dominates. The linear part is evidently the driving force behind the confining potential between the $q$ and $\bar{q}$.
\begin{figure}[h!]
\centering
\includegraphics[width=2.7in,height=1.9in]{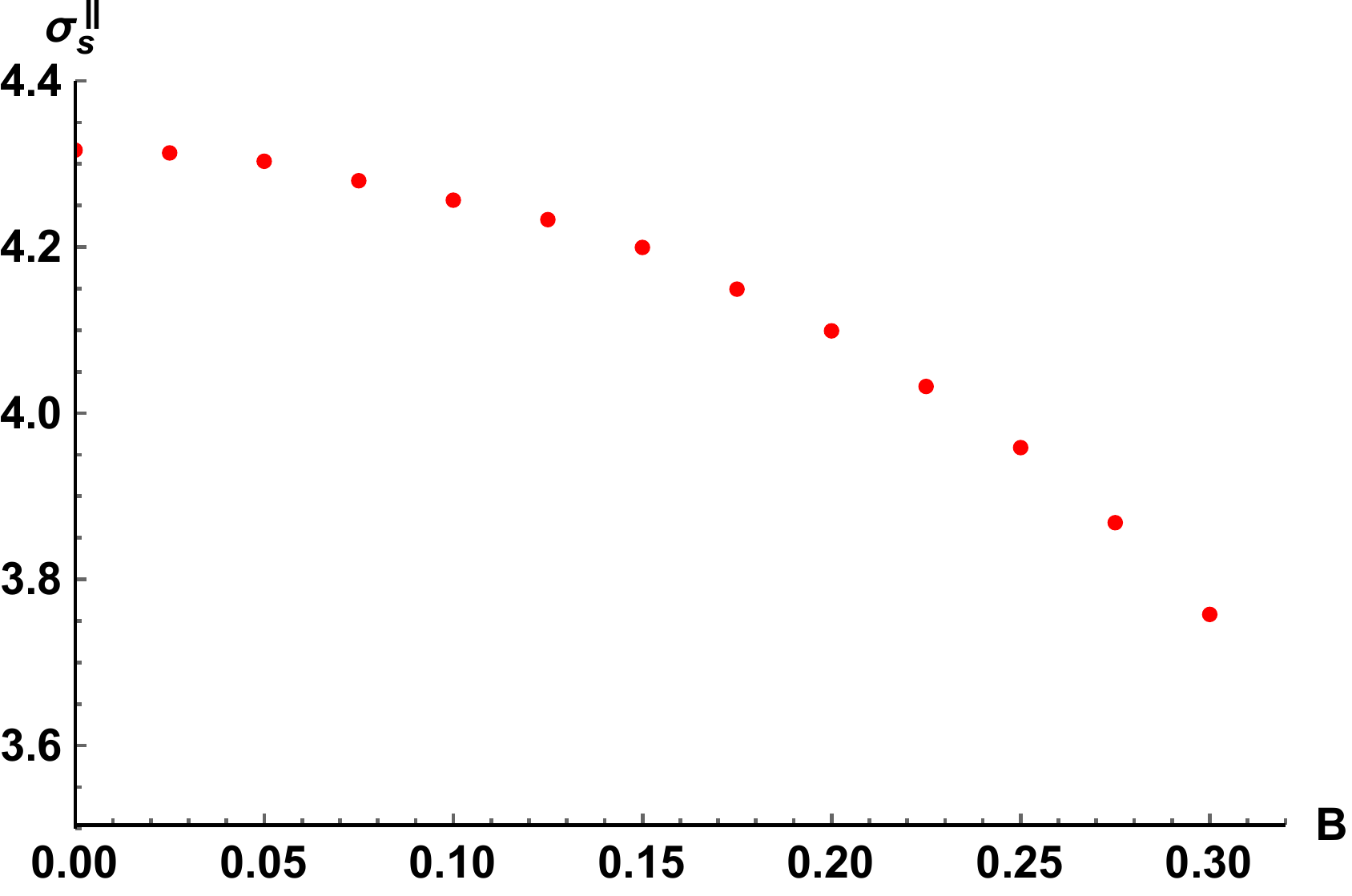}
\caption{ \small  $\sigma_{s}^{\|}$ as a function of $B$ in the thermal AdS background with $\mu=0$. In units \text{GeV}.}
\label{sigmapara}
\end{figure}
From the linear regime, where ${\cal {F}}^{\|}_{con} \propto \sigma_{s}^{\|}\ell^{\|}$, we can obtain the QCD string tension via $\sigma_{s}^{\|}=\frac{d {\cal {F}}^{\|}_{con}}{d \ell^{\|}}$. The behavior of the QCD string tension for a parallel orientation in terms of magnetic field for the thermal-AdS background (still $\mu=0$) is shown in Figure \ref{sigmapara}. We clearly observe a decreasing $\sigma_{s}^{\|}$, i.e.~a weaker confinement along the applied magnetic field. This is compatible with the lattice results that were reported in \cite{Bonati:2014ksa,Bonati:2016kxj}.

It is important to verify that we work in the regime of $\ell^{\|}$ and magnetic field for which the energy ${\cal {F}}^{\|}_{con}$ is actually lower than ${\cal {F}}^{\|}_{discon}$. For $B=0$, this happens to be the case because the integral of ${\cal {F}}^{\|}_{discon}$ is divergent \cite{Dudal:2017max}. Indeed, the upper limit of the integral in ${\cal {F}}^{\|}_{discon}$ is $z_h\to\infty$ and the integrand blows up for large $z$ if $B=0$. However in some cases with $B>0$, various form factors which appear in the disconnected integrand cause an exponential dampening in ${\cal {F}}^{\|}_{discon}$ at large $z$, allowing for a richer dynamics. With non-zero $B$, ${\cal {F}}^{\|}_{discon}$ can be lower than ${\cal {F}}^{\|}_{con}$ at larger values of $\ell$ and so would be the favoured string configuration.

\begin{figure}[h!]
\begin{minipage}[b]{0.5\linewidth}
\centering
\includegraphics[width=2.8in,height=2.0in]{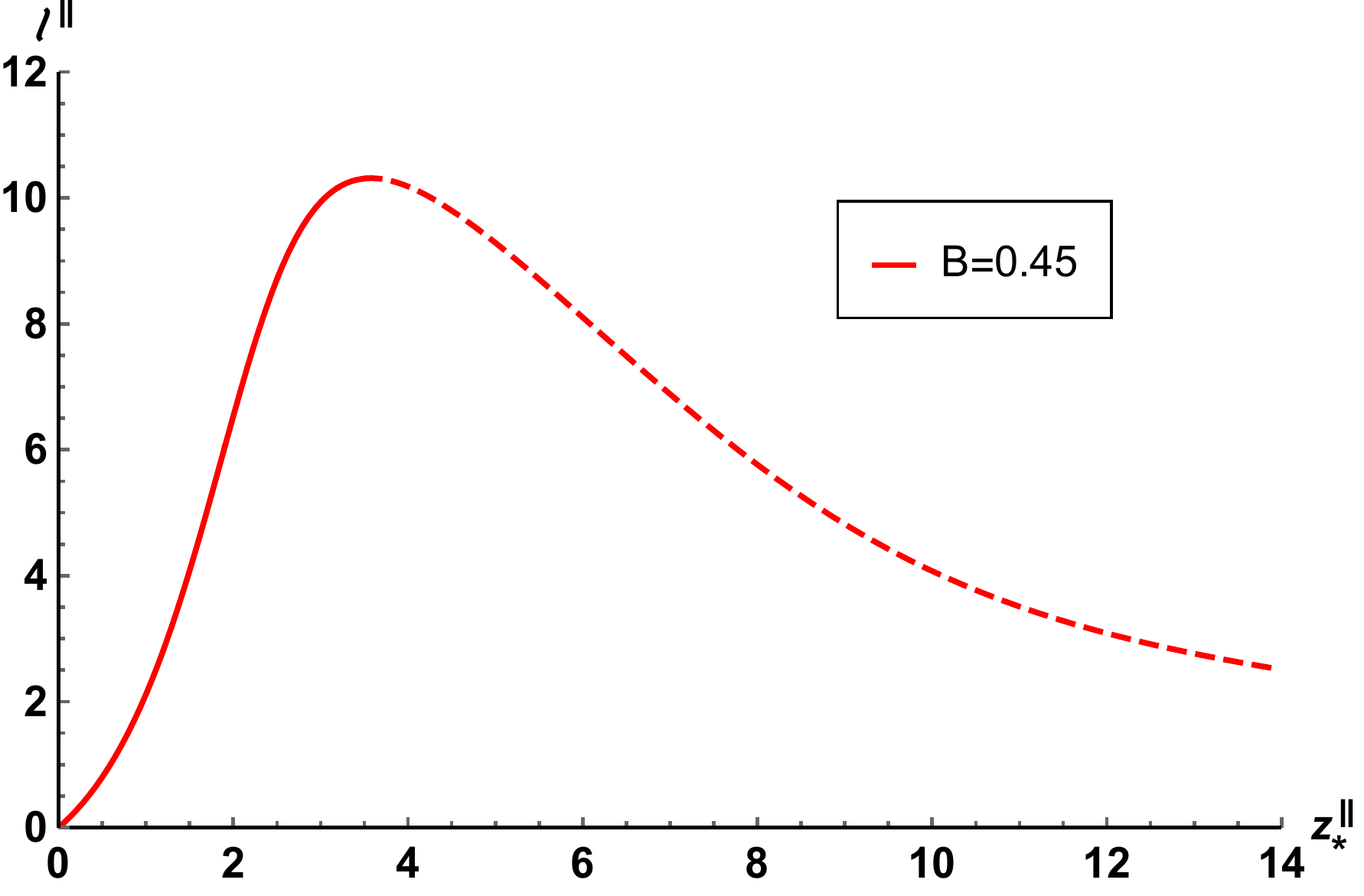}
\caption{ \small $\ell^{\|}$ as a function of $z^{\|}_*$ in the thermal AdS background and $\mu=0$. In units \text{GeV}.}
\label{ellparalzstar2}
\end{minipage}
\hspace{0.4cm}
\begin{minipage}[b]{0.5\linewidth}
\centering
\includegraphics[width=2.8in,height=2.3in]{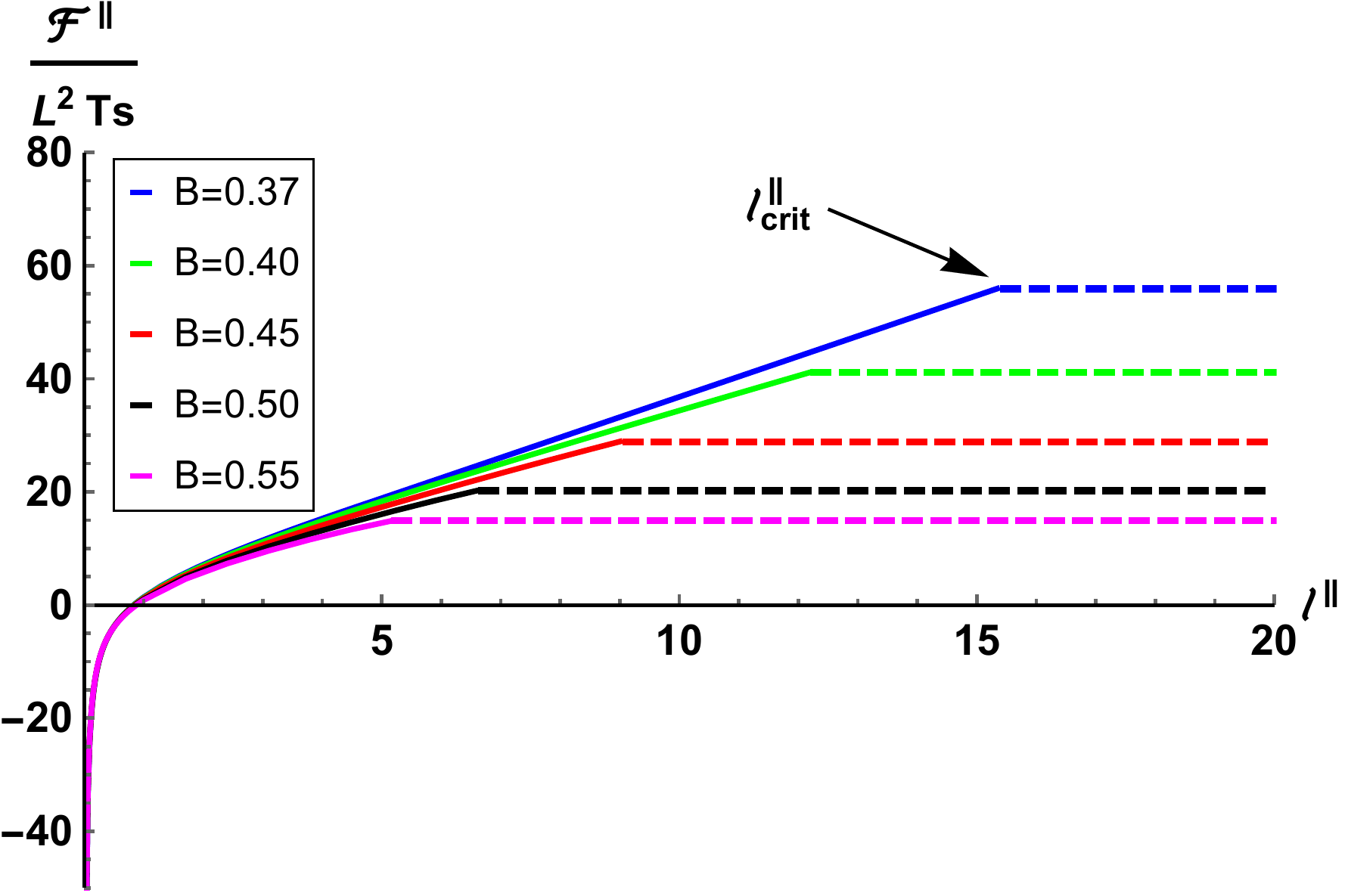}
\caption{\small ${\cal {F}}^{\|}$ as a function of $\ell^{\|}$ in the thermal AdS background for different (large) magnetic fields and $\mu=0$. In units \text{GeV}.}
\label{deltaf2}
\end{minipage}
\end{figure}

In particular, when we work with larger magnetic field values, viz.~approximately $B > 0.37~\text{GeV}$, the imaginary wall that appeared for smaller magnetic field values (approximately $B< 0.30~\text{GeV}$), disappears. For example, in Figure \ref{ellparalzstar2}, $\ell^{\|}$ as a function of $z_*^{\|}$ for $B=0.45~\text{GeV}$ is shown. We see that there are now two solutions of $z_*^{\|}$ for each value of $\ell^{\|}$ (provided $\ell<\ell_{max}$) and interestingly if one chooses the values of $z_*^{\|}$ before the maximum (smaller $z_*^{\|}$, red solid line part), one might still extract the linear behaviour for the potential in terms of $\ell$.

Via $\Delta{ \cal {F}}^{\|}={\cal {F}}^{\|}_{con} - {\cal {F}}^{\|}_{discon}$, we found that for large magnetic field there is a critical length of interquark distance, $\ell_{crit}^{\|}$, so that for $\ell^{\|}<\ell_{crit}^{\|}$ the $\Delta{ \cal {F}}^{\|}$ is negative and hence the connected configuration is favoured, whilst for $\ell^{\|}>\ell_{crit}^{\|}$ the $\Delta{ \cal {F}}^{\|}$ is positive, and so we must take into account the disconnected string configuration. Moreover, we find that if we increase the magnetic field further, the value of this $\ell_{crit}^{\|}$ decreases (see Figure~\ref{deltaf2}). Also, since ${\cal {F}}^{\|}_{discon}$ is actually independent of $\ell$, there would be no linear behavior for the potential either, i.e.~no more confinement. It means that even though ${\cal {F}}^{\|}_{con}$ exhibits the area law, the correct dynamics of $q, \bar{q}$ pair is actually described by the disconnected configuration, and the QCD string tension is zero.

For completeness, we depicted the behavior of ${\cal {F}}^{\|}$ as a function of $\ell^{\|}$ in the thermal AdS background for different (large) magnetic fields in Figure \ref{deltaf2}. Notice that for $\ell^{\|}<\ell_{crit}^{\|}$ the potential is linear suggesting confinement whereas for $\ell^{\|}>\ell_{crit}^{\|}$ it becomes independent of $\ell$ suggesting $q, \bar{q}$ pair breaking. Interestingly, if we focus on the linear parts of the potential in Figure \ref{deltaf2}, we see that the slope of these decreases when we increase the magnetic field. Naively, this suggests that even for large magnetic fields, in the linear regime, the parallel QCD string tension decreases for increasing magnetic field.

Although we do not have dynamical (light) quarks in the game, the behaviour shown in Figure \ref{deltaf2} resembles that of a string breaking when the energy stored in the string (flux tube) connecting the heavy $q, \bar q$ gets large enough to support pair creation, i.e.~what would happen in genuine QCD \cite{Bali:2005fu}.
\subsubsection{Perpendicular case}
Let us now investigate the effect of $B$ on the QCD string tension when it is perpendicular to the quark-antiquark distance $\ell^{\perp}$. Our embedding is then different from the parallel case. We again choose the static gauge, i.e.~$\tau= t$ and $\sigma= y_{2}$, to parameterize the two-dimensional string world-sheet. Analogously as for the parallel case from eqs. (\ref{nambugoto2}) and (\ref{metric5}), we can obtain connected and disconnected solutions that minimize the Nambu-Goto action.

The connected solution is still a $\cup$-shape configuration,
\begin{eqnarray}
{\cal {F}}^{\perp}_{con}= \frac{L^2}{\pi \ell ^{2}_{s}} \int_{\epsilon}^{z^{\perp}_*} dz \frac{z^{\perp 2}_*}{z^2}\frac{\sqrt{e^{B^2 z^2} g(z)}~e^{2A_{s}(z)-2A_{s}(z^{\perp}_*)}}{\sqrt{e^{B^2 z^2} g(z) z^{\perp 4}_*  e^{-4A_{s}(z^{\perp}_*)} - e^{B^2 z^{\perp 2}_*}g(z^{\perp}_*) z^4 e^{-4A_{s}(z)}}}~,
\label{paraconn}
\end{eqnarray}
where $z^{\perp}_*$ is the turning point for the perpendicular case and $\epsilon$ again the UV cut-off. The relation between the $q, \bar q$ separation $\ell^{\perp}$ and the $z^{\perp}_*$ is now
\begin{eqnarray}
\ell^{\perp} =2 \int_{\epsilon}^{z^{\perp}_*} dz \sqrt{\frac{ e^{B^2 z^{\perp 2}_*} g(z^{\perp}_*)}{e^{B^2 z^2} g(z)}}\frac{z^2~e^{-2A_{s}(z)}}{\sqrt{e^{B^2 z^2} g(z) z^{\perp 4}_*  e^{-4A_{s}(z^{\perp}_*)} - e^{B^2 z^{\perp 2}_*}g(z^{\perp}_*) z^4 e^{-4A_{s}(z)}}}~.
\label{ellconn}
\end{eqnarray}
In this case, the free energy of the disconnected solution reads
\begin{eqnarray}
{\cal {F}}^{\perp}_{discon}= \frac{L^2}{\pi \ell ^{2}_{s}} \int_{\epsilon}^{z_h} dz \frac{e^{2A_{s}(z)}}{z^2}
\label{paraconn}
\end{eqnarray}
where $z_h \to \infty$ for the thermal AdS background. The employed renomalization scheme for ${\cal {F}}^{\perp}_{con}$ and ${\cal {F}}^{\perp}_{discon}$ is similar to the parallel case.

\begin{figure}[h!]
\begin{minipage}[b]{0.5\linewidth}
\centering
\includegraphics[width=2.7in,height=1.9in]{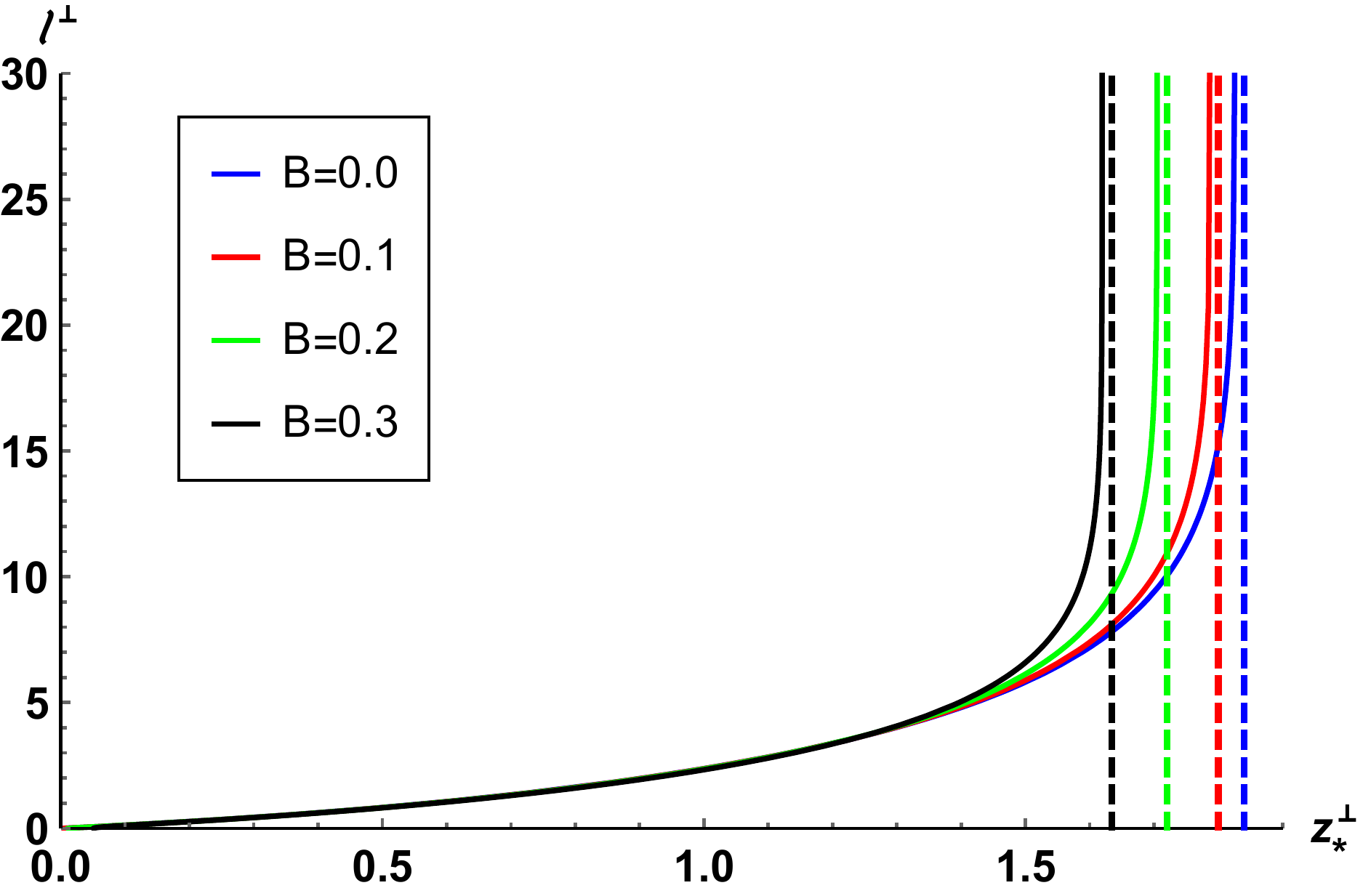}
\caption{ \small $\ell^{\perp}$ as a function of $z^{\perp}_*$ in the thermal AdS background for different (small) magnetic fields and $\mu=0$. In units \text{GeV}.}
\label{ellperpzstar}
\end{minipage}
\hspace{0.4cm}
\begin{minipage}[b]{0.5\linewidth}
\centering
\includegraphics[width=2.8in,height=2.3in]{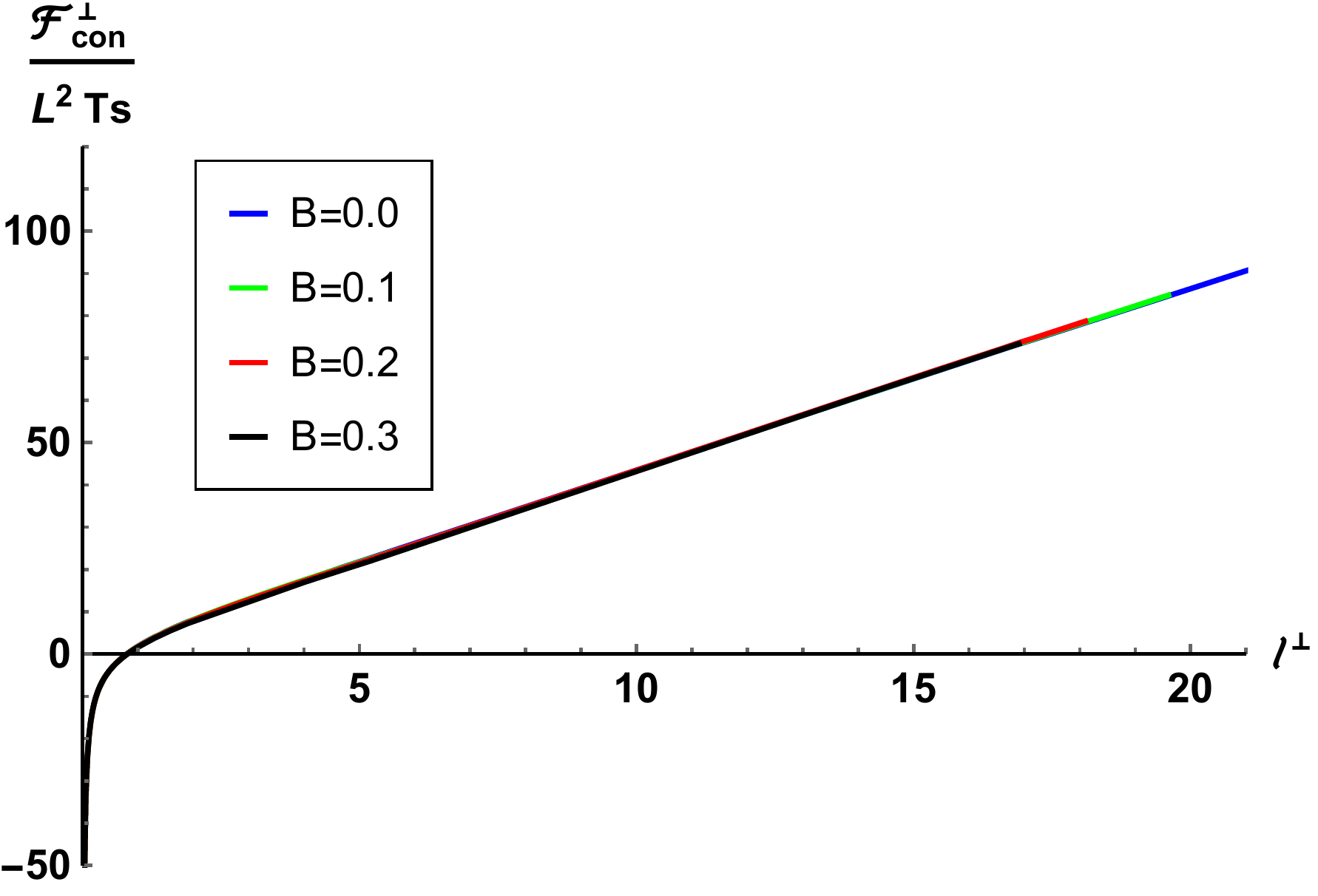}
\caption{ \small  ${\cal {F}}^{\perp}_{con}$ as a function of $\ell^{\perp}$ in the thermal AdS background for different (small) magnetic fields and $\mu=0$. In units \text{GeV}.}
\label{fconn2}
\end{minipage}
\end{figure}

\begin{figure}[h!]
\begin{minipage}[b]{0.5\linewidth}
\centering
\includegraphics[width=2.7in,height=1.9in]{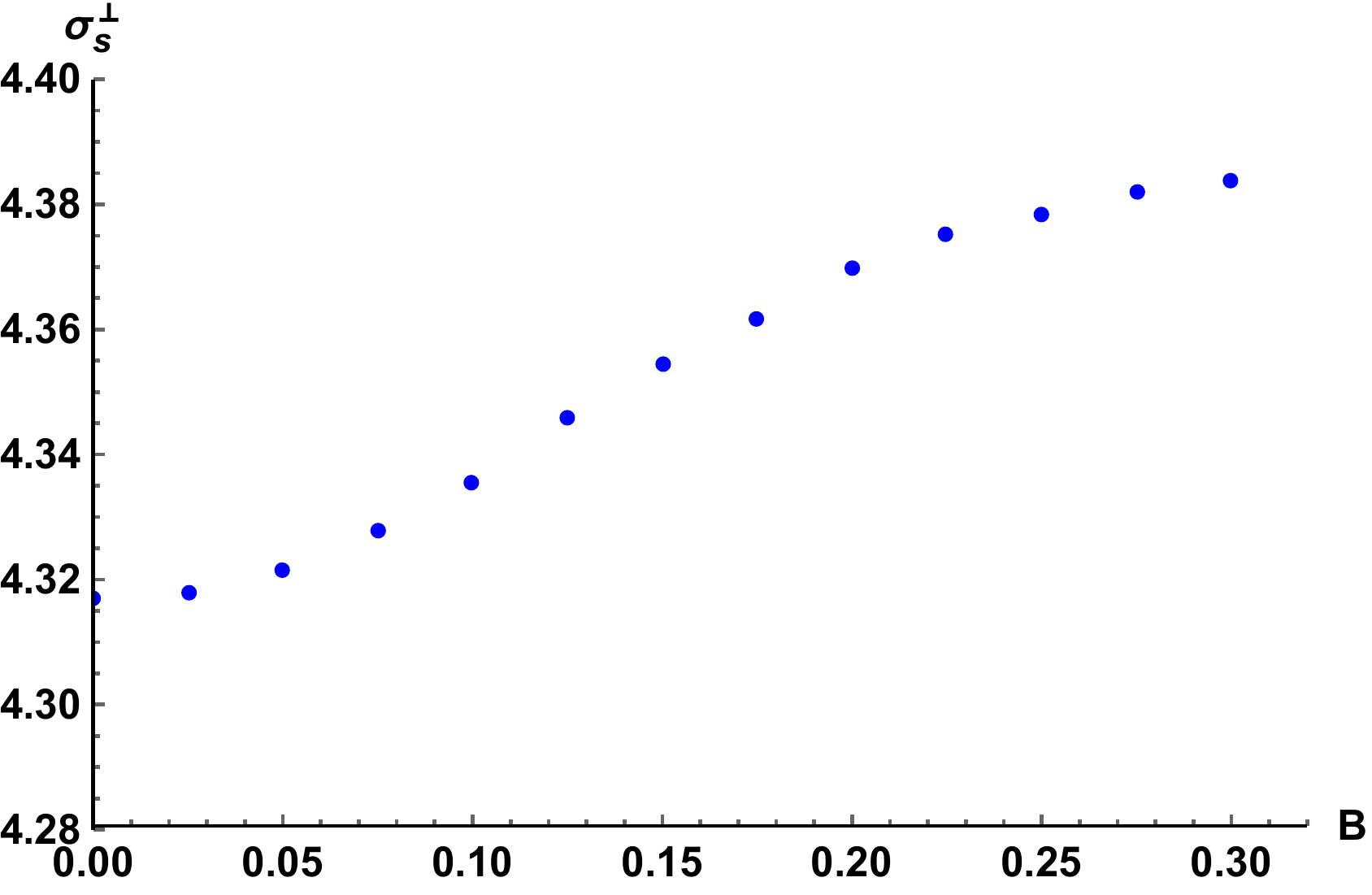}
\caption{ \small $\sigma_{s}^{\|}$ as a function of $B$ in the thermal AdS background with $\mu=0$. In units \text{GeV}.}
\label{sigmaperp}
\end{minipage}
\hspace{0.4cm}
\begin{minipage}[b]{0.5\linewidth}
\centering
\includegraphics[width=2.8in,height=2.3in]{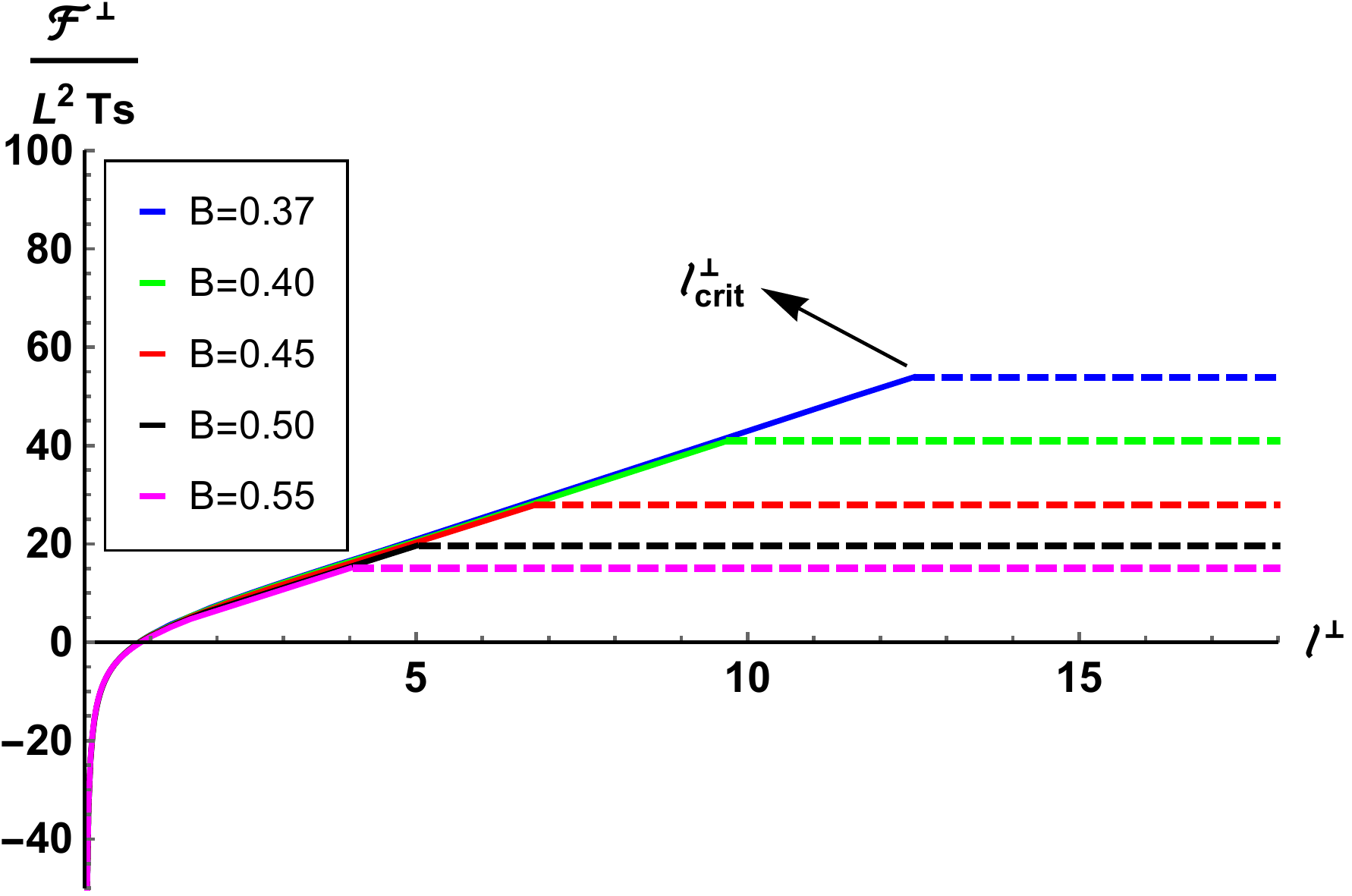}
\caption{ \small  ${\cal {F}}^{\perp}$ as a function of $\ell^{\perp}$ in the thermal AdS background for different (large) magnetic fields and $\mu=0$. In units \text{GeV}.}
\label{deltaf5}
\end{minipage}
\end{figure}

First, we consider $\ell^{\perp}$ vs. $z^{\perp}_*$ behaviour for different values of $B$ with $\mu=0$. This is shown in Figure \ref{ellperpzstar}. Similar to the parallel case, we again encounter the ``imaginary wall'' that captures the confinement on the gauge theory side.
Interestingly, the location of the imaginary wall now shifts toward lower values of $z$ by increasing $B$ in the perpendicular case, thereby suggesting less penetration of the string world sheet for higher $B$. This is different from the parallel case where the ``imaginary wall'' shifts to higher values of $z$. The corresponding $q, \bar{q}$ connected ${\cal {F}}^{\perp}_{con}$ free energy behaviour for different (small) $B$ is shown in Figure \ref{fconn2}. Each of them can now again be fitted with the Cornell potential \cite{Eichten:1974af,Eichten:1978tg},
\begin{eqnarray}
\frac{{\cal {F}}^{\perp}_{con}}{L^2 T_{s}}= -\frac{\kappa^\perp}{\ell^{\perp}}+ \sigma_{s}^{\perp} \ell^{\perp} + C^\perp~.
\label{cornell2}
\end{eqnarray}
We obtain the QCD string tension in the perpendicular case via $\sigma_{s}^{\perp}=\frac{d {\cal {F}}^{\perp}_{con}}{d \ell^{\perp}}$ by focussing on the linear regime of Figure \ref{fconn2}. We find that the QCD string tension (shown in Figure \ref{sigmaperp}) in the perpendicular case (slightly) increases with $B$. This is again in contrast with the parallel case where the string tension decreases with $B$. Similar as in the parallel case, here our maximal choice for the small magnetic field is $B\leq0.30~\text{GeV}$. This enhanced perpendicular confinement is also compatible with the lattice results of \cite{Bonati:2014ksa,Bonati:2016kxj}.

In the perpendicular case as well, a critical length appears with (large) magnetic field such that ${\cal {F}}^{\perp}_{con}<{\cal {F}}^{\perp}_{discon}$ for $\ell^{\perp}< \ell_{crit}^{\perp}$ whereas ${\cal {F}}^{\perp}_{con}>{\cal {F}}^{\perp}_{discon}$ for $\ell^{\perp}> \ell_{crit}^{\perp}$. Accordingly, we have plotted ${\cal {F}}^{\perp}$ as a function of $\ell^{\perp}$ for different (large) magnetic fields in Figure \ref{deltaf5}.  The connected string configuration which is relevant for $\ell^{\perp}< \ell_{crit}^{\perp}$ is denoted by a solid line whereas the disconnected string configuration which is relevant for $\ell^{\perp}> \ell_{crit}^{\perp}$ is denoted by a flat dashed line. This behaviour follows that of the earlier discussed parallel case.

For completeness, we also extracted estimates for both $\kappa^\|$ and $\kappa^\perp$ in terms of $B$, finding up to very good accuracy that $\kappa^\perp(B)\approx \kappa^\|(B) \approx\kappa(B=0)$, thereby suggesting that the Coulomb strength is barely $B$-dependent. This is in line with the (extrapolated) lattice estimates of \cite{Bonati:2016kxj}. On the other hand, the constant terms in the Cornell-fitted potential are affected by $B$ and are different for parallel and perpendicular cases, while a lattice extrapolation suggested this constant to be universal as well.

\section{Outlook}
We constructed a sensible, magnetic field dependent gravity dual of QCD with the interesting features of anisotropic confinement expressed by an orientation dependent string tension, next to inverse magnetic catalysis for the deconfinement sector.

In a next phase of research, we should study whether the inverse catalysis phenomenon extends to the chiral sector by adding a probe scalar degree of freedom to the theory that describes the chiral condensate, following earlier works like \cite{Erlich:2005qh,Colangelo:2011sr,Dudal:2015wfn}. Available lattice  data suggests that the chiral and deconfinement transition continues to coincide even in presence of a magnetic field, this by using various dedicated order parameters \cite{Bali:2011qj}. It is a priori not clear if this will also hold holographically, see for example \cite{Gursoy:2017wzz}.

Moreover, our model could also be fruitful to study, now in a gravitationally consistent setting, the melting and transport properties of charmonia in magnetic fields, thereby improving upon \cite{Dudal:2014jfa,Dudal:2015kza,Dudal:2018rki,Braga:2018zlu,Braga:2019yeh}, see also \cite{Suzuki:2016fof,Yoshida:2016xgm,Bonati:2015dka,Sadofyev:2015hxa,Cho:2014loa,Iwasaki:2018czv}.

It would also be interesting to find out to what extent the observation of \cite{Gursoy:2017wzz} that the inverse catalysis turns into catalysis again if the chemical potential gets larger, is generically valid.  In our case, this would only happen at the level of the chiral transition, since we confirmed already the inverse catalysis in presence of any chemical potential.

Another interesting direction to extend our work will be to use the entanglement structure, in particular the entanglement entropy, of holographic QCD phases to investigate (inverse) magnetic catalysis, following works like \cite{Dudal:2016joz,Knaute:2017lll,Gursoy:2018ydr}.

Moreover, we can further improve our bottom-up model to mimic QCD to the best extent possible. One open question is whether we can find a gravity solution that remains valid up to (much) larger values of the magnetic field, to further probe the lattice predictions of \cite{Bonati:2016kxj}, which reported via an extrapolation, the destruction of the parallel string tension for sufficiently large magnetic field. Next to that, we can also adapt a form factor to match the running of the QCD strong coupling constant in the IR as well.

\section*{Acknowledgments}
A.H.~would like to thank a scholarship that has been awarded by the Ministry of Science, Research and Technology (Department of Scholarship and Students' s Affairs Abroad) of the Islamic Republic of Iran which made this research possible. The work of S.M.~is supported by the Department of Science and Technology, Government of India under the Grant Agreement number IFA17-PH207 (INSPIRE Faculty Award).

\appendix
\section{(In)dependence of the potential on temperature, magnetic field or chemical potential}
The following figures illustrate the almost independence of the dilaton potential $V(z)$ on the parameter $z_h$ (or $T$), $B$ and $\mu$.

\begin{figure}[h!]
\centering
\includegraphics[width=2.8in,height=2.3in]{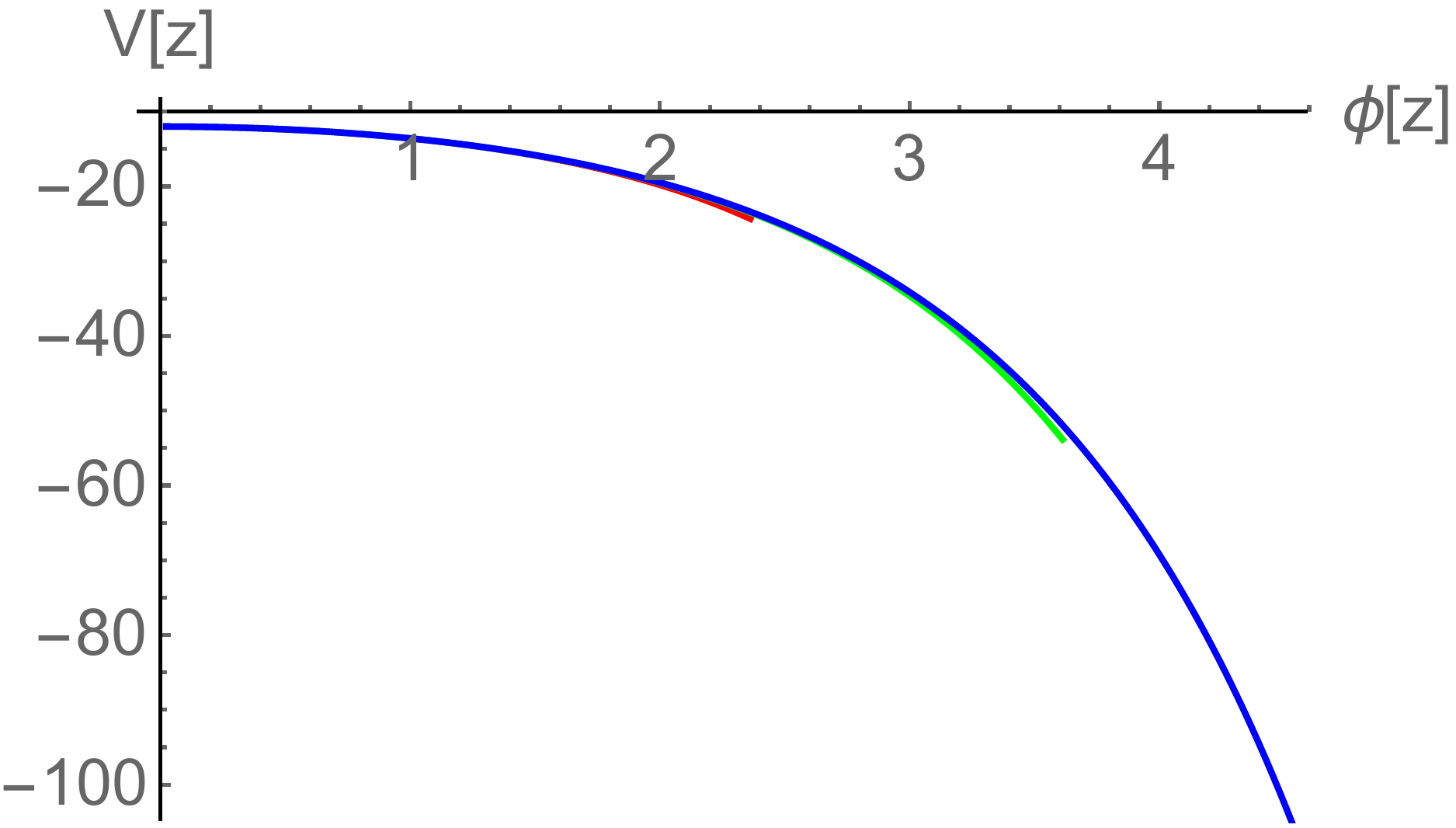}
\caption{ \small The variation of potential as a function of $z$ for different $z_h$. Here $\mu=0$ and $B=0$ are considered. Here red, green and blue curves correspond to $z_h=1$, $z_h=1.5$ and $z_h=2$ respectively.}
\label{zvsVz}
\end{figure}
\begin{figure}[h!]
\begin{minipage}[b]{0.5\linewidth}
\centering
\includegraphics[width=2.8in,height=2.3in]{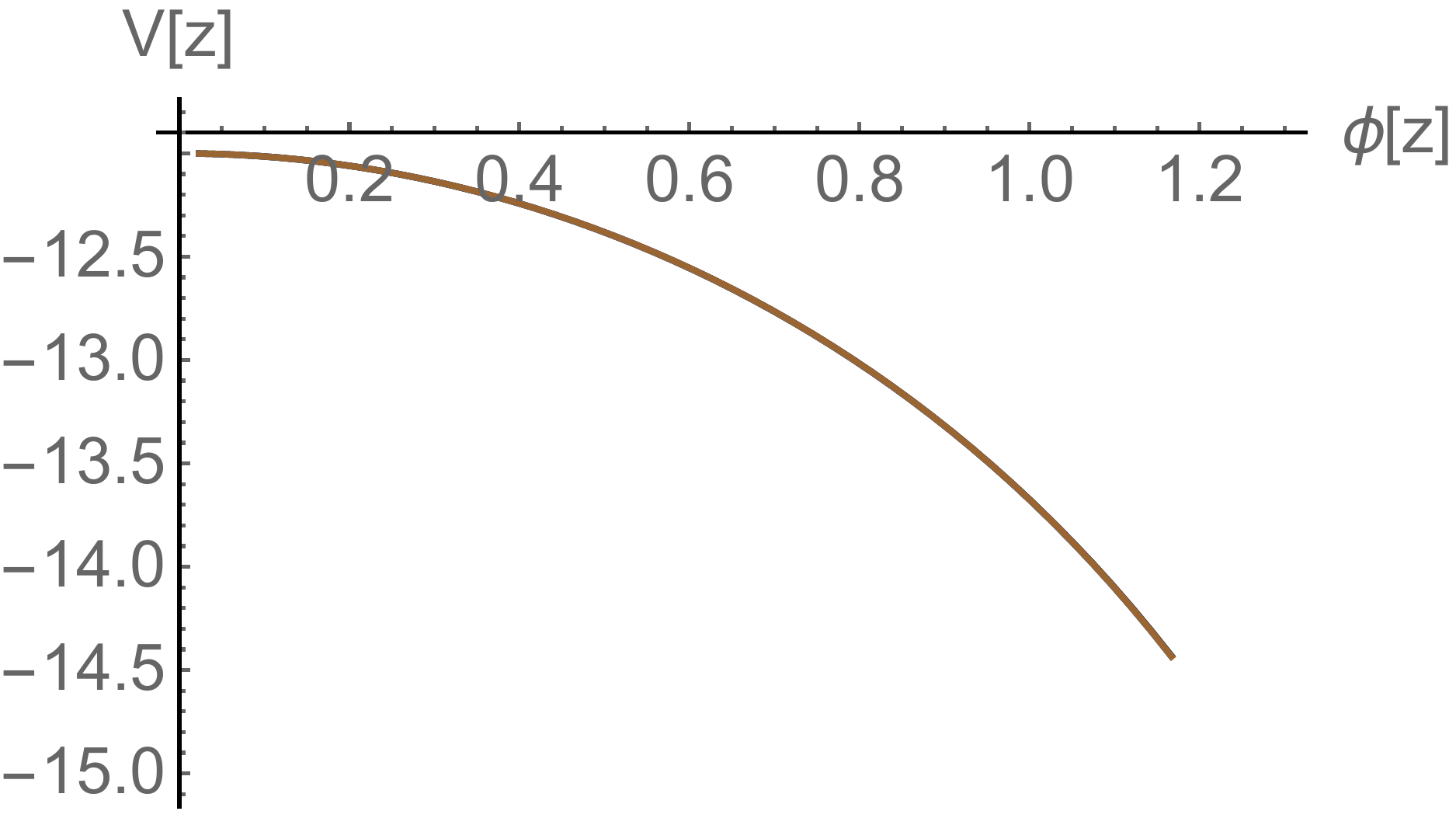}
\caption{ \small The variation of potential with different values of $\mu$. Here $B=0$ and $z_h=1.5$ are considered. Red, green, blue and brown curves correspond to $\mu=0$, $0.2$, $0.4$ and $0.6$ respectively.}
\label{ZvsVvsMuB0}
\end{minipage}
\hspace{0.4cm}
\begin{minipage}[b]{0.5\linewidth}
\centering
\includegraphics[width=2.8in,height=2.3in]{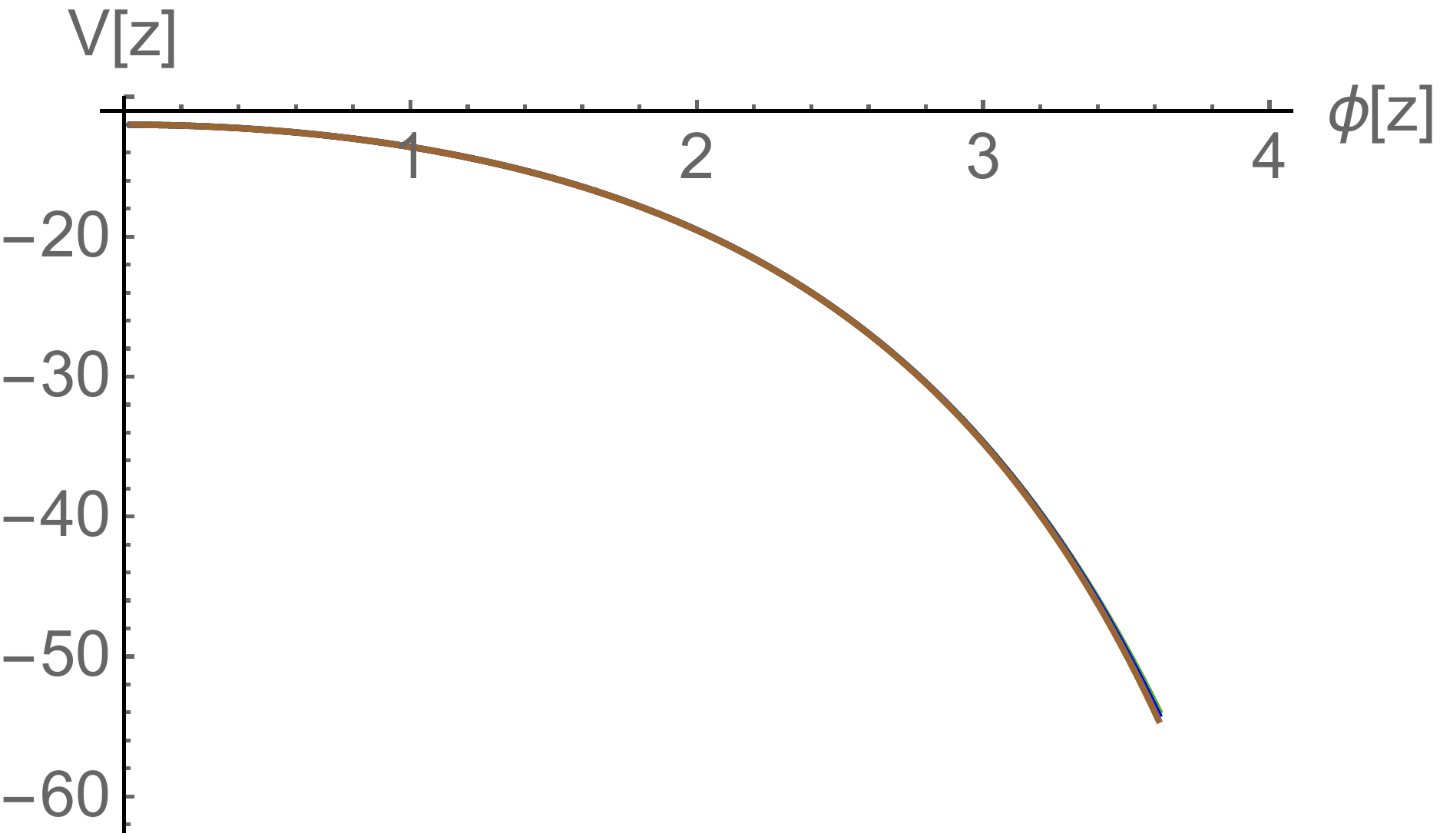}
\caption{\small The variation of potential with different values of $\mu$. Here $B=0$ and $z_h=0.5$ are considered. Red, green, blue and brown curves correspond to $\mu=0$, $0.2$, $0.4$ and $0.6$ respectively. }
\label{ZvsVvsMuB0LargeBH}
\end{minipage}
\end{figure}
\begin{figure}[h!]
\begin{minipage}[b]{0.5\linewidth}
\centering
\includegraphics[width=2.8in,height=2.3in]{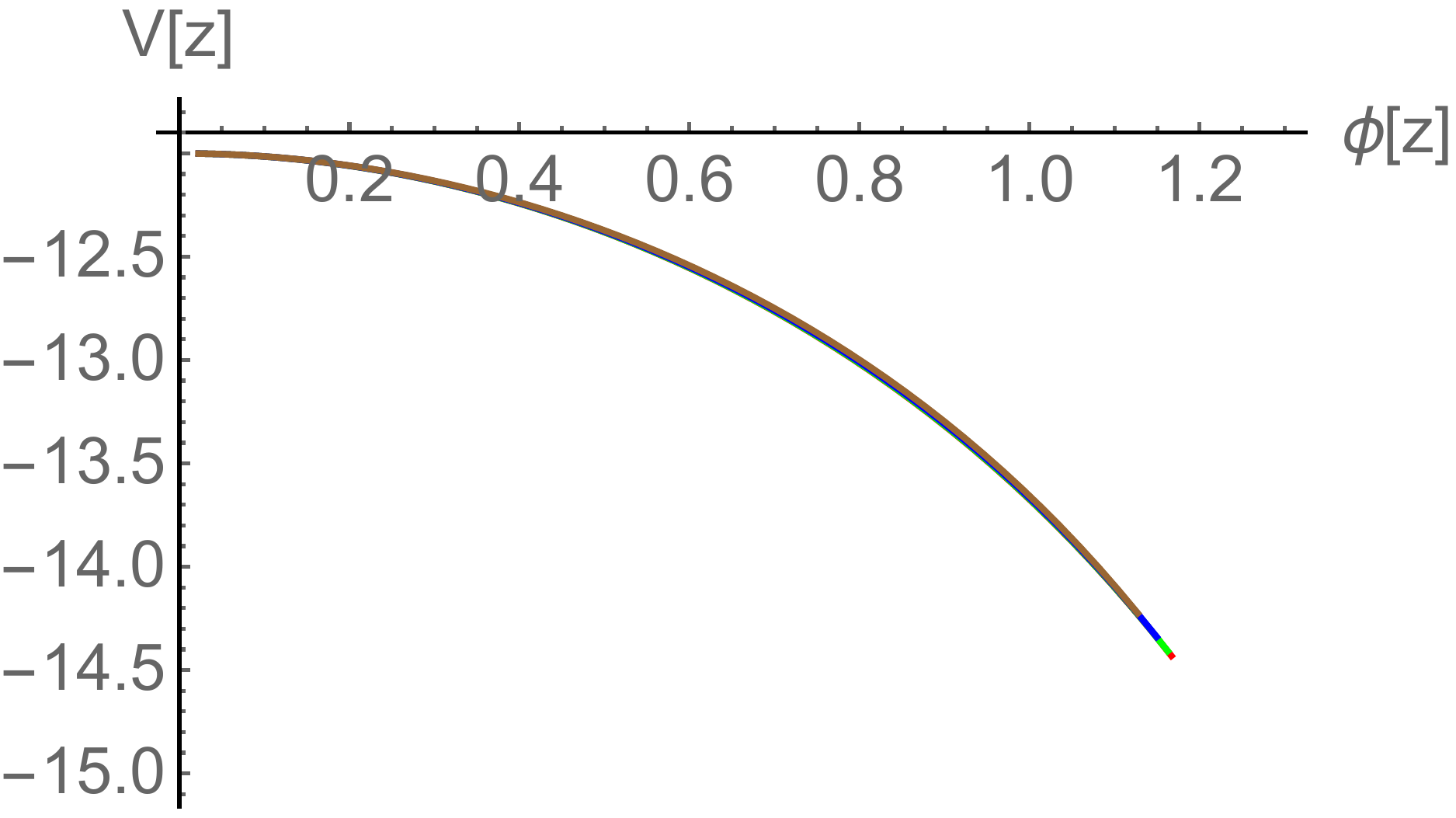}
\caption{ \small The variation of potential with different values of $B$. Here $\mu=0$ and $z_h=1.5$ are considered. Red, green, blue and brown curves correspond to $B=0$, $0.1$, $0.2$ and $0.3$ respectively.}
\label{ZvsVvsBMu0}
\end{minipage}
\hspace{0.4cm}
\begin{minipage}[b]{0.5\linewidth}
\centering
\includegraphics[width=2.8in,height=2.3in]{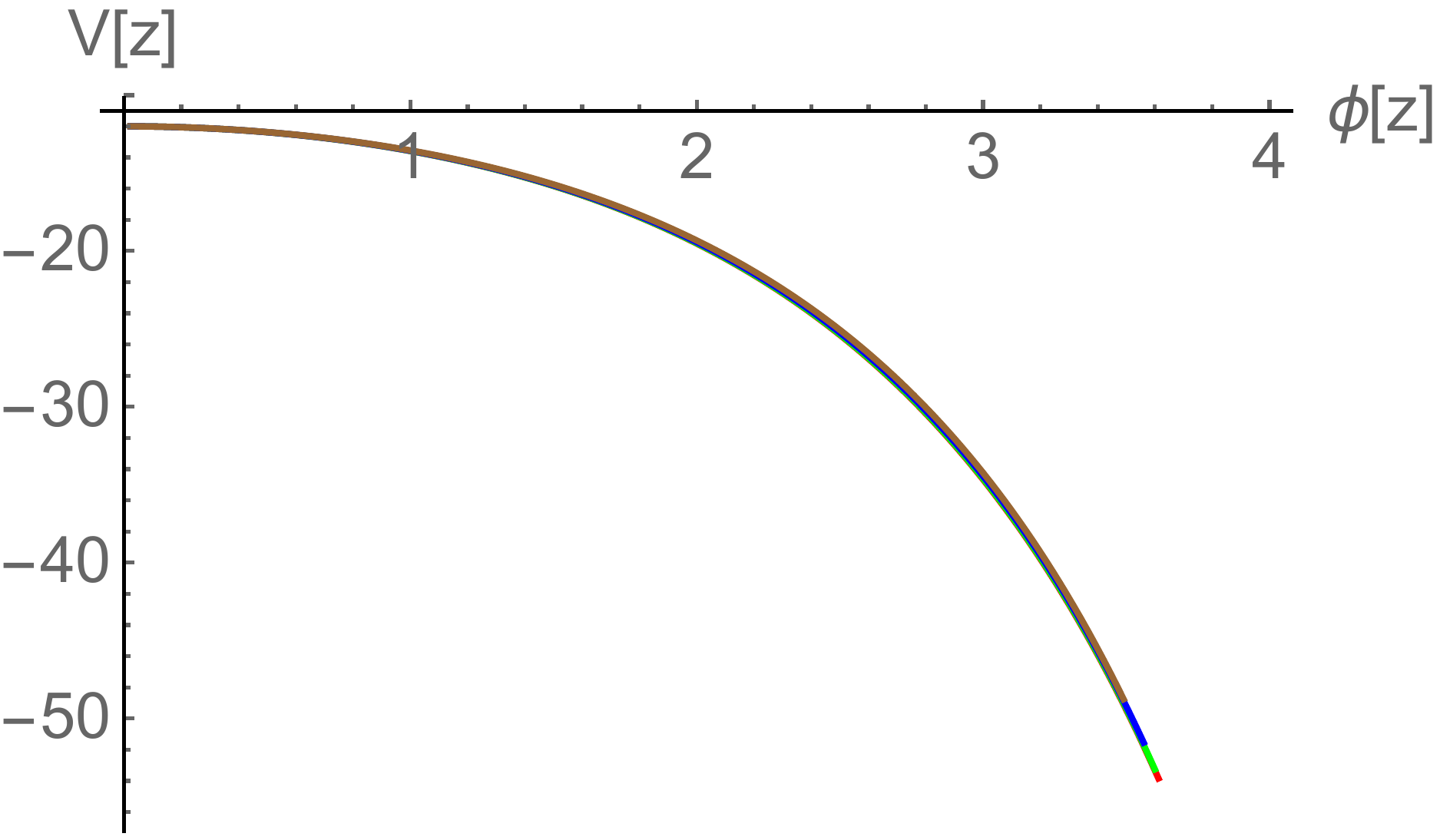}
\caption{\small The variation of potential with different values of $B$. Here $\mu=0$ and $z_h=1.5$ are considered. Red, green, blue and brown curves correspond to $B=0$, $0.1$, $0.2$ and $0.3$ respectively.}
\label{ZvsVvsBMu0Large}
\end{minipage}
\end{figure}

\end{document}